\documentclass[english]{emulateapj}

\usepackage{graphicx}
\usepackage{apjfonts}
\def\gs{\mathrel{\raise1.16pt\hbox{$>$}\kern-7.0pt %
\lower3.06pt\hbox{{$\scriptstyle \sim$}}}}         %
\def\ls{\mathrel{\raise1.16pt\hbox{$<$}\kern-7.0pt %
\lower3.06pt\hbox{{$\scriptstyle \sim$}}}}         %
 \newcommand{\ltsimeq}{\raisebox{-0.6ex}{$\,\stackrel 
        {\raisebox{-.2ex}{$\textstyle <$}}{\sim}\,$}} 
\newcommand{\gtsimeq}{\raisebox{-0.6ex}{$\,\stackrel 
        {\raisebox{-.2ex}{$\textstyle >$}}{\sim}\,$}}

\slugcomment{}
\shorttitle{Optical Line Diagnostics of Optically faint ULIRGs}
\shortauthors{}

\usepackage{babel}
\makeatother
\begin{document}

\title{Optical Line Diagnostics of $z\approx$2 Optically Faint Ultra-Luminous Infrared Galaxies in the $Spitzer$ Bo\"otes Survey}

\author{K. Brand\altaffilmark{1}, A. Dey\altaffilmark{2}, V. Desai\altaffilmark{3}, B.~T. Soifer\altaffilmark{3,4}, C. Bian\altaffilmark{3}, L. Armus\altaffilmark{4}, M.~J.~I. Brown\altaffilmark{5}, E. Le Floc'h\altaffilmark{6}, S.~J. Higdon\altaffilmark{7,8}, J. ~R. Houck\altaffilmark{8}, B.~T. Jannuzi\altaffilmark{2}, D.~W. Weedman\altaffilmark{8}} 

\altaffiltext{1}{Space Telescope Science Institute, 3700 San Martin Drive, Baltimore, MD 21218; brand@stsci.edu}
\altaffiltext{2}{National Optical Astronomy Observatory, 950 North Cherry Avenue, Tucson, AZ 85726} 
\altaffiltext{3}{Division of Physics, Mathematics and Astronomy, California Institute of Technology, 320-47, Pasadena, CA 91125}
\altaffiltext{4}{Spitzer Science Center, California Institute of Technology, 220-6, Pasadena, CA 91125}
\altaffiltext{5}{Department of Astrophysical Sciences, Princeton University, Peyton Hall, Princeton, NJ 08544-1001}
\altaffiltext{6}{Steward Observatory, University of Arizona, Tucson, AZ, 85721}
\altaffiltext{7}{Physics Dept. P. O. Box 803, Georgia Southern University, Statesboro, GA 30460}
\altaffiltext{8}{Astronomy Department, Cornell University, Ithica, NY 14853}

\begin{abstract}
We present near-infrared spectroscopic observations for a sample of ten optically faint luminous infrared galaxies ($R-[24]\ge 14$) using Keck NIRSPEC and Gemini NIRI. The sample is selected from a 24 $\rm \mu m$ $Spitzer$ MIPS imaging survey of the NDWFS Bo\"otes field. We measure accurate redshifts in the range $1.3\ltsimeq z\ltsimeq3.4$. Based on either emission line widths or line diagnostics, we find that all ten galaxies harbor luminous AGN. Seven  sources are type I AGN, exhibiting broad ($>$1900 km s$^{-1}$) H$\alpha$ or H$\beta$ emission lines; the remaining three are type II AGN. Given their large mid-IR luminosities and faint optical magnitudes, we might expect these sources to be heavily extincted quasars, and therefore only visible as type II AGN. The visibility of broad lines in 70\% of the sources suggests that it is unlikely that these AGN are being viewed through the mid-plane of a dusty torus. For four of the sources we constrain the H$\alpha$/H$\beta$ Balmer decrement and estimate the extinction to the emission line region to be large for both type I and type II AGN, with $A_{\rm H\alpha}\gtsimeq 2.4-5$~mag. Since the narrow-line region is also extincted and the UV continuum emission from the host galaxies is extremely faint, this suggests that much of the obscuration is contributed by dust on large ($\sim$kpc) scales within the host galaxies. These sources may be examples of "host-obscured" AGN which could have space densities comparable or greater to that of optically luminous type I AGN with similar bolometric luminosities.  
\end{abstract}


\section{Introduction}

Recent results from the $Spitzer~Space~Telescope$ \citep{wer04} have shown that luminous infrared galaxies (LIRGs) become increasingly important at high redshift \citep{lef04}. They dominate the energy density of the Universe at $z>1$ and their extrapolated infrared luminosities suggest that they are galaxies undergoing a highly active phase in the growth of their bulges and / or accompanying super-massive black holes (SMBHs). Although their power source appears to be primarily due to starbursts at lower luminosities ($\rm L_{IR}(8-1000~\mu m)~<~10^{12}L_\odot$), AGN activity becomes increasingly important at higher luminosities (\citealt{vei95}; \citealt{dud99}; \citealt{bra06}). 

We have uncovered a population of sources which are bright at 24 $\rm \mu m$ but optically very faint (Dey et al. in prep.). Although only a relatively small fraction of the total 24 $\rm \mu m$ population, the fraction increases from 5\% at 24 $\rm \mu m$ flux densities (f$_{24}$) greater than 1 mJy to 15\% at f$_{24} >$ 0.3 mJy (Dey et al. in preparation), suggesting that they become increasingly important either at high redshifts or low luminosities.  Hereafter, we will refer to sources with $R-[24]>14$ as optically faint luminous infrared galaxies. Optically faint luminous infrared galaxies have optical to mid-infrared colors which are more than a magnitude redder than Arp 220 (an extreme dusty source in the local Universe). Their huge bolometric luminosities imply that they must be undergoing rapid growth via dust-enshrouded AGN and/or starburst activity. However, because these sources are also optically faint, they are also likely very dust-obscured and will have been largely missed by previous optical surveys. Determining the evolution of their luminosity function and the relative contribution of AGN and starburst activity to their bolometric luminosity is critical in understanding how the most massive galaxies and their SMBHs built up.

Although optically faint luminous infrared galaxies are difficult to study at optical wavelengths, their extreme red SEDs makes them good candidates for mid-infrared (mid-IR) spectroscopy. Our  follow-up of a sub-sample of 58 of these sources with $Spitzer$ / IRS, yielded redshifts in the range $z\approx 2-3$  for 34 of the sources, based primarily on the strong 9.7 $\rm \mu m$ silicate absorption feature (\citealt{hou05}; \citealt{wee06}; Higdon et al. in preparation; see also \citealt{yan05}). An additional 9 have probable redshifts derived from weak silicate absorption features \citep{wee06}. 15 sources have featureless power-law IRS spectra \citep{wee06} from which one can obtain little further information. It is not clear whether this is due to the sources lying at redshifts for which the silicate absorption feature falls out of the observable band ($z>2.5$), or because they have only weak silicate absorption features \citep{wee06}. In these extreme sources, the IRS spectra often have low signal-to-noise ratios and can be difficult to interpret. Usually, the silicate absorption feature is the only strong observable feature. The strength of this feature is determined primarily by the geometry \citep{lev06} or clumpiness \citep{spo06} of the dust region and can be strong in both AGN- and starburst-dominated sources (although \citealt{spo06} show that starburst-dominated sources tend to have stronger PAH features and only shallow absorption features). The silicate strength may simply provide us with a lower limit to the line of sight obscuration towards the nucleus.  Redshifts and diagnostics from other wavelengths will help us understand these sources better and may be able to identify starburst signatures that may be overwhelmed by the hot dust continuum in the mid-IR. 

Rest-frame optical spectra can potentially help in understanding the nature of these sources. One can determine redshifts for the sources with featureless power-law spectra in the mid-IR, and confirm and refine the redshift measurements for those sources with broad silicate absorption features. By determining the widths and relative strengths of the optical lines, one can also help constrain the properties of the energy source. H$\alpha$ is a particularly useful line in this regard because it is generally luminous and it can be significantly broadened by the presence of an AGN. In the cases in which the broad-line region is obscured from view, one can use the line ratios of [O III] $\lambda$5007 / H$\beta$ and [NII] $\lambda$6583 / H$\alpha$ to distinguish between narrow-line AGN and H II region-like galaxies (e.g., \citealt{bal81}; \citealt{vei87}; \citealt{vei95}; \citealt{kau03}; \citealt{kew06}). The $R$-[24]$>$14 selection criteria means our sample should be particularly affected by dust extinction. H$\alpha$ should be less attenuated than emission lines at shorter wavelengths and the Balmer decrement (i.e., the H$\alpha$ to H$\beta$ ratio) provides a useful measure of the optical extinction.

For the redshifts of the sources in our sample, the typically strong, diagnostically useful optical lines of H$\alpha$ and H$\beta$ are shifted into the near-IR. Being very red, optically faint luminous infrared galaxies can also be relatively bright in the near-IR ($K\approx 19-20$) compared to the optical ($R>24$). Near-IR spectroscopy using the largest telescopes presents the best (and in many cases, the only) chance of determining the nature of these extreme sources. In this paper, we present and analyze Keck NIRSPEC and Gemini NIRI near-IR spectra of 10 high redshift optically faint ULIRGs. We use optical emission line diagnostics to determine their redshifts and provide insights into their primary power source. 

The paper is structured in the following way. The sample selection criteria are presented in Section~2. In Section~3, we discuss the observational techniques, data reduction, and calibration. The near-IR spectra are presented in Section~4. We also discuss each source individually. In Section~5, we discuss the characteristics of the sample. Section~6 is a discussion section in which we investigate the broader implications of our results. A cosmology of $H_0 = 70 {\rm ~km ~s^{-1} ~Mpc^{-1}}$, $\Omega_M$=0.3, and $\Omega_\Lambda$=0.7 is assumed throughout. 

\section{The Sample}

Our sample of 10 optically faint ULIRGs is selected from a 24 $\rm \mu m$ survey of the NOAO Deep Wide-Field Survey (NDWFS; \citealt{jan99}) Bo\"otes field.  The NDWFS Bo\"otes field comprises deep optical ($R\ltsimeq$25.5) and near-IR ($K\ltsimeq$18.6) imaging. A thorough description of the NDWFS survey will be provided in Jannuzi et al. and Dey et al. (both in preparation). The $\sim$9.3 deg$^2$ region has been mapped by $Spitzer$  MIPS at 24 $\rm \mu m$ (f$_{24}$) and comprises $\approx$20,000 sources down to a 5$~\sigma$ depth of $\approx$0.3 mJy (Le Floc'h et al. in preparation). For more details regarding the creation of the 24$\rm \mu m$ catalog, see \citet{bro06} and \citet{bra06}. The $\sim$2,500 $R-[24]>14$ sources will be described in more detail in Dey et al. (in preparation). 

Our sample is selected from the $R-[24]>14$ sources using one of the following three criteria: 

\begin{itemize}
\item[(1)]{Sources with a known redshift in the range $1.25<z<1.6$, $2.1<z<2.6$, or $3.2<z<3.7$. At these redshifts, the strongest expected emission lines (i.e., H$\alpha$ / [NII] and H$\beta$ / [OIII] ) have rest-frame wavelengths which are observable from the ground. Although this sample is biased to sources for which we could obtain redshifts (i.e., sources with strong UV emission lines observable with optical spectroscopy and/or a strong silicate absorption feature within the $Spitzer$ IRS wave-band), it provided us with objects for which the near-IR spectra would yield detections or useful constraints on the rest-frame optical lines. This sample consists of 2 sources for which we obtained optical redshifts (Desai et al. in preparation) but which exhibited featureless power-law spectra in the mid-IR (SST24 J143424.4+334543 and SST24 J142644.3+333051), 1 source with an optical redshift with no IRS observation (SST24 J143011.3+343439), and 3 sources for which the silicate feature in the IRS spectra provided an estimate of the redshift (SST24 J142827.1+354127, SST24 J143312.7+342011, and SST24 J143028.5+343221).}

\item[(2)]{Sources observed by $Spitzer$ IRS which yielded only featureless mid-IR spectra and no redshift measurement. Observing these sources with near-IR spectroscopy may be the only way to obtain a redshift and determine their nature. Four such sources were targeted by our near-IR spectroscopy, but only two had detected emission lines and are presented here (SST24 J142842.9+342409 and SST24 J142939.1+353558).}

\item[(3)]{Sources which are bright in the $K$-band ($K\approx17.5$-$18.0)$. Six such sources were selected and targeted by our Gemini NIRI observations but only two sources exhibited strong emission lines and are presented here (SST24 J143027.1+344007 and SST24 J142800.7+350455).}

\end{itemize}
  
In all cases without a bright $K$-band detection, we preferentially chose sources that had a $K<18$ star within 20$^{\prime\prime}$. This made target acquisition quicker and easier and provided a reference from which to measure where spectrum of the target should fall (see section \ref{sec:obs}). We also chose sources with the brightest detections in the $K$-band. Although the $K$-band light may not be correlated with the line emission strength, detecting continuum is useful for confirming that the source is in the slit and for providing a relative flux calibration of the source. The $K$-band magnitudes were obtained from either our NDWFS ONIS observations, the FLAMEX survey \citep{els06}, or follow-up Keck NIRC observations \citep{wee06}. 

A table summarizing our sample and their auxiliary data is provided in Table~\ref{tab:sample}. 

\begin{deluxetable*}{llllll}
\tabletypesize{\scriptsize}
\setlength{\tabcolsep}{0.005in}
\tablecolumns{6} 
\tablewidth{0pc}
\tablecaption{\label{tab:sample} Observed and derived properties of the sample of 10 optically faint ULIRGs.} 
\tablehead{ 
\colhead{Name\tablenotemark{a}} & \colhead{J142827.1+354127} & \colhead{J143027.1+344007} & \colhead{J143312.7+342011} & \colhead{J143011.3+343439\tablenotemark{b}} & \colhead{J143028.5+343221}\\
}
\startdata  
RA (J2000)\tablenotemark{c}     & 14:28:27.191		& 14:30:27.203		& 14:33:12.708		& 14:30:11.367		& 14:30:28.521 \\
Dec (J2000)\tablenotemark{c}   	& 35:41:27.70  	& 34:40:07.86 		& 34:20:11.20   	& 34:34:39.68 		& 34:32:21.34  \\
$B_W$ mag (Vega)\tablenotemark{d}& 23.85		& 25.63		& 24.62		& 22.88 		& 25.17		\\
$R$ mag (Vega)\tablenotemark{d}	& 22.78 		& 24.76		& 24.37		& 22.02 		& 24.45		\\
$I$  mag (Vega)\tablenotemark{d}& 21.64 		& 23.19		& 23.57		& 21.84		& 24.16		\\
f$_{24\rm \mu m}$ (mJy)		& 10.55		& 1.17			& 1.76			& 1.12			& 1.27		\\
$R$-[24]			& 15.7			& 15.3			& 15.3			& 12.5			& 15.1		\\
IRS $z$\tablenotemark{e}	& 1.33			& $-$			& 2.2			& $-$ 			& 2.2			\\
Optical $z$\tablenotemark{e,f}	& 1.292		& $-$			& $-$ 			& 2.12			& 2.176 		\\
\tableline
Near-IR instrument / band	& NIRSPEC / $H$	& NIRI / $J$/$H$	& NIRSPEC / $K$ 	&NIRSPEC / $K$/$H$      &NIRSPEC / $K$\\
Exp (s)				& 3600			& 3360 / 3840		& 3600			&3600/3600 		& 3600		\\
Date (UT)			& 2005may26 		& 2006jun08/07	        & 2006may11		&2005may27		&2005may27	\\
\tableline
Near-IR $z$			& 1.292$\pm$0.001 / 1.383$\pm$0.001& 1.370$\pm$0.001& 2.114$\pm$0.001& 2.114$\pm$0.001& 2.178$\pm$0.001\\
Continuum amplitude\tablenotemark{g}& 1.10$\pm$0.00	& 0.59$\pm$0.01	& 0.21$\pm$0.04	& 0.23$\pm$0.05	& 0.36$\pm$0.05 \\
H$\alpha~\lambda$ (\AA)		& 15045 / 15689	& 15557		& 20434		& 20436 		& 20857\\
H$\alpha$ flux\tablenotemark{h}	&19.5$\pm$5.3	/115.7$\pm$12.3& 390.7$\pm$12.8 & 375.1$\pm$51.8 	& 101.3$\pm$13.1	& 104.5$\pm$18.3\\
H$\alpha~\sigma$\tablenotemark{n}(\AA)& 4.8$\pm$0.9/ 86.8$\pm$5.2& 42.8$\pm$0.7       & 90.8$\pm$8.3    	& 13.6$\pm$1.1	        & 16.0$\pm$1.5\\
L(H$\alpha$)\tablenotemark{k}($\rm10^{42} erg~s^{-1})$&0.19$\pm$0.04/1.3$\pm$0.1& 4.4$\pm$0.1& 12.4$\pm$1.7 & 3.4$\pm$0.4	&3.7$\pm$0.6\\
FWHM(H$\alpha$) (km s$^{-1}$)    &227$\pm$42/3919$\pm$233  & 1940$\pm$34	& 3138$\pm$288	        & 469$\pm$37		&542$\pm$49\\
EW(H$\alpha$) (rest) (\AA)	&7.7$\pm$2.1/44.1$\pm$4.7	  & 280$\pm$10	&583$\pm$136		& 142$\pm$38		&90$\pm$20\\
$\rm[NII]$$\lambda$\tablenotemark{i}(\AA)&15092 / 15736 & 15606		& 20499		& 20501		& 20922 \\
$\rm [NII]$ flux\tablenotemark{h}&$<$ 98.8/$<$ 64.8	& $<$ 56.2		& 5.2$\pm$13.3	        & 33.2$\pm$17.6         & 50.5$\pm$13.5\\
$\rm [NII]$ $\sigma$\tablenotemark{n}(\AA)& $-$ /$ -$	& $-$			& 5.6$\pm$7.2		&13.5$\pm$3.5 	        &14.4$\pm$2.2\\
$\rm [NII]$ / H$\alpha$		& $<$ 5.1/($<$ 0.57) 	& ($<$ 0.14)		& (0.01$\pm$0.04) 	& 0.33$\pm$0.18	&0.48$\pm$0.15\\
Continuum amplitude\tablenotemark{g} & $-$		& 0.19$\pm$0.01	& $-$			& 0.14$\pm$0.02	& $-$\\
H$\beta~\lambda$ (\AA)		& $-$			& 11521		& $-$			&15138 		& $-$\\
H$\beta$ flux\tablenotemark{h}	& $-$			& $<$17.6		& $-$			&$<$ 11.9		& $-$\\
H$\beta~\sigma$\tablenotemark{n}(\AA)& $-$ 		& $-$			& $-$			& $-$			& $-$\\
L(H$\beta$)\tablenotemark{k}($\rm10^{42} erg~s^{-1})$& $-$& $-$			& $-$ 			& $-$			& $-$\\
FWHM(H$\beta$) (km s$^{-1}$)& $-$			& $-$			& $-$			& $-$			& $-$\\
EW(H$\beta$) (rest)		& $-$			& $-$			& $-$			& $-$			& $-$\\
$\rm[OIII]$$~\lambda$\tablenotemark{j}(\AA)& $-$	& 11866		& $-$			&15591			& $-$\\
$\rm[OIII]$ flux\tablenotemark{h}& $-$	        	& 2.8$\pm$2.8		& $-$			&37.6$\pm$10.8	& $-$\\
$\rm[OIII]$$~\sigma$\tablenotemark{n}(\AA)& $-$	& 6.3$\pm$3.7		& $-$			&22.5$\pm$3.5		& $-$\\
L([OIII])\tablenotemark{k}($\rm10^{42} erg~s^{-1})$& $-$	& 0.03$\pm$0.03	& $-$			&1.2$\pm$0.3 		& $-$\\
FWHM([OIII]) (km s$^{-1}$)	& $-$			& 375$\pm$220  	& $-$			&(1020$\pm$160)	& $-$\\
$\rm [OIII]$ / H$\beta$		& $-$			& ($>$ 0.6)		& $-$			&$>$ 3.1		& $-$\\
H$\alpha$ / H$\beta$		& $-$			& $>$ 22.5		& $-$			& $>$ 8.5		& $-$ \\
E(B-V)				& $-$ 			& $>$1.93		& $-$			& $>$ 1.00		& $-$  \\
A(H$\alpha$)			& $-$ 			& $>$ 4.64		& $-$			& $>$ 2.40		& $-$\\
class	\tablenotemark{l}	&SB/AGN1		& AGN1			& AGN1			&AGN2			& AGN2\\
$\rm L_{IR} (10^{13} L_{\odot}$)\tablenotemark{m}& 2.8	& 0.3			& 1.3			& 0.8			& 1.0\\
IRS spectrum			& \citealt{des06}	& $-$ 			&Higdon et al. in prep.& $-$			&Higdon et al. in prep.\\
IRS features			& Si Abs		& $-$			& Si Abs		& $-$			&Deep Si Abs	\\
\enddata
\end{deluxetable*} 

\addtocounter{table}{-1}

\begin{deluxetable*}{llllll}
\tabletypesize{\scriptsize}
\setlength{\tabcolsep}{0.002in}
\tablecolumns{6} 
\tablewidth{0pc} 
\tablecaption{\label{tab:sample} Continued.} 
\tablehead{ 
\colhead{Name$^a$} & \colhead{J142842.9+342409} & \colhead{J142800.7+350455} & \colhead{J143424.4+334543} & \colhead{J142939.1+353558} & \colhead{J142644.3+333051}\\
}
\startdata  
RA (J2000)\tablenotemark{c}     & 14:28:42.913		& 14:28:00.724		& 14:34:24.500 	& 14:29:39.159		& 14:26:44.330\\
Dec (J2000)\tablenotemark{c}    &  34:24:09.44 	& 35:04:55.31		& 33:45:43.40		&  35:35:58.24		&  33:30:52.06 \\
$B_W$ mag (Vega)\tablenotemark{d}& 23.57		& 25.00		& (24.57)		& (27.37)		& $-$		\\
$R$ mag (Vega)\tablenotemark{d} &  22.54		& 24.28		& 25.62		& (26.90) 		& 24.34		\\
$I$  mag (Vega)\tablenotemark{d}&  21.47		& 23.89		& (24.30)		& (25.78)		& $-$		\\
f$_{24\rm \mu m}$ (mJy)          & 3.11			& 0.92			& 0.86			& 1.16			& 1.14		\\
$R$-[24]		       & 14.1			& 14.5			& 15.6			& 17.4			& 14.8		\\
IRS $z$\tablenotemark{e}	& ?			& $-$			& ?			& ?   			& ?			\\
Optical $z$\tablenotemark{e,f}	& ?			& $-$			&  2.26  		& ? 			& 3.355 		\\
\tableline
Near-IR instrument / band	&NIRSPEC / $K$/$H$      & NIRI  / $H$/$K$ 	&  NIRSPEC / $K$ 	&  NIRSPEC / $K$ 	&  NIRSPEC / $K$\\
Exp (s)				& 3600 / 7200		& 3840 / 960		& 3600			&3600 			& 3600		\\
Date (UT)			& 2006may10/11 	& 2006jun08/jul29	& 2006may11		&2006may11		&2006may10	\\
\tableline
Near-IR $z$			& 2.180$\pm$0.001	& 2.223$\pm$0.001	& 2.263$\pm$0.001 	& 2.498$\pm$0.001	& 3.312$\pm$0.001 \\
Continuum amplitude\tablenotemark{g}& 0.32$\pm$0.07	& 0.00$\pm$0.01	& 0.30$\pm$0.03 	& 0.10$\pm$0.03	& $-$		 \\
H$\alpha~\lambda$ (\AA)		& 20870		& 21152		& 21411		&  22957 		& $-$		\\
H$\alpha$ flux\tablenotemark{h}	&673.2$\pm$41.9	& 230.3$\pm$27.8	& 27.0$\pm$11.8	& 200.0$\pm$31.1	& $-$		\\
H$\alpha~\sigma$\tablenotemark{n}(\AA)& 93.6$\pm$3.2  	& 114.9$\pm$6.9	& 21.0$\pm$4.6 	& 67.1$\pm$5.7	        & $-$		\\
L(H$\alpha$)\tablenotemark{k}($\rm10^{42} erg~s^{-1})$& 24.0$\pm$1.5& 8.6$\pm$1.0& 1.1$\pm$0.5		& 9.9$\pm$1.5		&$-$	\\
FWHM(H$\alpha$) (km s$^{-1}$)    & 3167$\pm$110	        & 3833$\pm$229	        & 692$\pm$153	        & 2063$\pm$176	        & -			\\
EW(H$\alpha$) (rest) (\AA)	& 672$\pm$154  	& $-$			& 27.7$\pm$12.5	& 564$\pm$189   	& -			\\
$\rm[NII]$$\lambda$\tablenotemark{i}(\AA)&20936	& 21219		& 21479 		& 23029		& -			 \\
$\rm [NII]$ flux\tablenotemark{h}&$<$ 103.7		& $<$ 62.1		& 124.9$\pm$16.1	& $<$158.3 		& -			\\
$\rm [NII]$ $\sigma$\tablenotemark{n}(\AA)& $-$ 	& $-$			& 31.9$\pm$2.6	        &- 			& -			\\
$\rm [NII]$ / H$\alpha$		& ($<$ 0.15)	 	& ($<$ 0.3)		&  4.6$\pm$2.1		& ($<$ 0.79)		& -			\\
Continuum amplitude\tablenotemark{g} & 0.01$\pm$0.01	& 0.11$\pm$0.01	& $-$			& $-$			& 1.07$\pm$0.02\\
H$\beta~\lambda$ (\AA)		& 15459		& 15669		& $-$			& $-$ 			& 20964\\
H$\beta$ flux\tablenotemark{h}	& $<$ 42.9 		& $<$ 59.8		& $-$			& $-$			& 57.9$\pm$17.7\\
H$\beta~\sigma$\tablenotemark{n}(\AA)& $-$ 		& $-$			& $-$ 			& $-$			& 63.4$\pm$11.8\\
L(H$\beta$)\tablenotemark{k}($\rm10^{42} erg~s^{-1})$& $-$& $-$			& $-$ 			& $-$			& 5.7$\pm$1.7\\
FWHM(H$\beta$) (km s$^{-1}$)    & $-$			& $-$			& $-$			& $-$			&2234$\pm$398 \\
EW(H$\beta$) (rest)		& $-$			& $-$			& $-$			& $-$			& 12.5$\pm$3.8 \\
$\rm[OIII]$$~\lambda$\tablenotemark{j}(\AA)& 15922	& 16138		& $-$			& $-$ 			& 21591 \\
$\rm[OIII]$ flux\tablenotemark{h}& 104.8$\pm$3.6	& 88.2$\pm$6.0	        & $-$			& $-$			& 108.2$\pm$10.5 \\
$\rm[OIII]$$~\sigma$\tablenotemark{n}(\AA)& 36.5$\pm$0.7& 13.4$\pm$0.4	        & $-$			& $-$			& 16.3$\pm$0.8\\
L([OIII])\tablenotemark{k}($\rm10^{42} erg~s^{-1})$& 3.7$\pm$0.1& 3.3$\pm$0.2	& $-$			& $-$			&10.6$\pm$1.0 \\
FWHM([OIII]) (km s$^{-1}$)	& 1617$\pm$31  	& 585$\pm$19		& $-$			& $-$			& 533$\pm$27\\
$\rm [OIII]$ / H$\beta$		& ($>$ 2.44)		& ($>$1.5)		& $-$			& $-$ 			& (1.87$\pm$0.6)\\
H$\alpha$ / H$\beta$		& $>$15.7		& $>$ 3.8		& $-$			& $-$			& $-$ \\
E(B-V)	 		     & $>$1.59			& $>$ 0.22		& $-$			& $-$			& $-$   \\
A(H$\alpha$)		     & $>$ 3.81		& $>$ 0.54		& $-$			& $-$			& $-$	\\
class	\tablenotemark{l}	& AGN1		     & AGN1			& AGN2			& AGN1			& AGN1\\
$\rm L_{IR} (10^{13} L_{\odot}$)\tablenotemark{m}& 2.5	& 0.8			& 0.8			& 1.3			& 2.6\\
IRS spectrum	        	& \citealt{wee06} (15) & $-$ 			& \citealt{wee06}  (9)&\citealt{wee06} (24)&\citealt{wee06} (5)\\
IRS features	        	& No Si Abs.		& $-$			& Possible Si Abs. 	& Possible Si Abs. 	& featureless \\
\enddata
\tablenotetext{a}{~Abbreviated from the full name of SST24 J142827.1+354127 etc.}
\tablenotetext{b}{~Note that SST24 J143011.3+343439 is not strictly an optically faint ULIRG by our definition of $R-[24]>14$.}
\tablenotetext{c}{~The quoted positions of our targets are measured directly from the optical images. In cases where the source is optically invisible, we used the MIPS position corrected to be on the NDWFS frame.}
\tablenotetext{d}{~In cases in which no catalog magnitude exists, we quote the magnitude obtained within a 4 \arcsec aperture in brackets.}
\tablenotetext{e}{~For both the IRS and optical redshifts, $-$ denotes that the object was not observed and ? denotes that a redshift could not be determined. The IRS redshifts are accurate to within 0.1.}
\tablenotetext{f}{~The optical redshifts were obtained on Keck LRIS and DEIMOS (Desai et al. in preparation).} 
\tablenotetext{g}{~The continuum amplitude is given in units of 1E-18 $\rm erg~s^{-1}~cm^{-2}~$\AA$^{-1}$ and refers to the fitted value in around the H$\alpha$/[NII] or H$\beta$/[OIII] lines.}
\tablenotetext{h}{~All flux densities are given in units of 1E-18 $\rm erg~s^{-1}~cm^{-2}$.}
\tablenotetext{i}{~All values refer to [NII] $\lambda$6583. The [NII] $\lambda$6548 $\sigma$ is fixed to be the same as that of [NII] $\lambda$6583 and the [NII] $\lambda$6548 flux is fixed to be 1/3 of [NII] $\lambda$6583.}
\tablenotetext{j}{~All values refer to [OIII] $\lambda$5007. The [OIII] $\lambda$4959 $\sigma$ is fixed to be the same as that of [OIII] $\lambda$5007 and the [OIII] $\lambda$4959 flux is fixed to be 1/3 of [OIII] $\lambda$5007.}
\tablenotetext{k}{~The luminosities are from the measured flux densities and do not include any corrections for the substantial correction due to slit losses (see Section~\ref{sec:flux}) or dust extinction (see Section~\ref{sec:dust}).} 
\tablenotetext{l}{~Sources are classified as AGN1 if they exhibit broad ($>$1000 km s$^{-1}$) Balmer lines and as AGN2 if they exhibit narrow Balmer lines and [NII]/H$\alpha$ and [OIII]/H$\beta$ ratios characteristic of AGN activity.}
\tablenotetext{m}{~The intrinsic infrared luminosity, L$\rm_{IR}(8-1000 \mu m)$, is estimated from L$_{24}$ using a bolometric correction of 7 (typical for AGN-dominated ULIRGs; \citealt{xu01})}
\tablenotetext{n}{~The Gaussian 1$\sigma$ width of the emission line (=2.263*FWHM)}
\end{deluxetable*} 

\section{Observations and Data Reduction}
\label{sec:obs}

Our sample was observed with either the Near Infrared Spectrograph (NIRSPEC; \citealt{mcl98}) on the Keck II telescope or the Near Infrared Imager (NIRI; \citealt{hod03}) on the Gemini North telescope. We discuss the observations separately in the following two sections and then discuss the data reduction which was similar for both data sets. 

\subsection{Keck NIRSPEC}

The majority (8/10) of our sample were observed with NIRSPEC on the Keck II telescope on 26th $-$ 27th May 2005 UT and 10th $-$ 11th May 2006 UT in photometric conditions. NIRSPEC is designed to operate over the wavelength region 0.94 to 5.4 $\rm \mu m$. We observed with both the $K$-band (NIRSPEC 7) and $H$-band (NIRSPEC 5) filters. The observing band and date for each individual object is given in Table~\ref{tab:sample}. Observations were obtained in low-resolution mode (R$\approx$1500) with a 0\farcs76 wide and 42$^{\prime\prime}$ long slit. Typical seeing was 0\farcs7 on the 26th $-$ 27th May 2005 UT and 0\farcs5 on the 10th $-$ 11th May 2006 UT.

Our sources are typically too faint to acquire with short integrations on the acquisition camera, SCAM. Instead, we took 2 minute exposures with the K$^\prime$ filter on SCAM, aligned the slit on a nearby ($<$20\arcsec away) brighter source and performed a blind offset to the target. In a few cases in which the optical images showed the target to be elongated, we used a specific position angle to obtain the maximum amount of light down the slit. In all other cases, we set the position angle such that the offset star was also in the slit. In cases in which the target was very faint, this helped to determine exactly where the target should fall on the array. For sky subtraction, the observations were split into 10 minute exposures, and the object was nodded along the slit by 10\arcsec-20\arcsec (taking care that the position of the brighter object used for acquisition did not coincide with any of the dither positions of the target). 

Four sources which were known to have $2.1 \le z \le 2.4$ from either the optical or IRS spectra were observed in the $K-$band to obtain measurements of the H$\alpha$ and [NII] emission lines.  Four sources with no redshift information were also observed in the $K-$band (we only present the results of the 2 sources which were detected; the remaining 2 sources are discussed in Section~\ref{sec:midir}). For 2 sources exhibiting strong H$\alpha$ emission, we also observed the sources in the $H-$band to determine line strengths and widths of the [OIII] $\lambda\lambda$5007, 4959 doublet and H$\beta$ lines. SST24 J142827.1+354127 is known to lie at $z=$1.292 and was observed in the $H-$band to target H$\alpha$ and [NII]. SST24 J142644.3+333051, known to have $z=$3.355 from optical spectroscopy, was targeted in the $K-$band to determine line strengths and widths of the [OIII] $\lambda\lambda$5007, 4959 doublet and H$\beta$ lines. Exposure times are listed in Table~\ref{tab:sample}; they were typically $\approx$1 hour per filter.

\subsection{Gemini / NIRI}

Six sources were observed with NIRI on the Gemini telescope on 7th $-$ 8th June 2006 UT, and 29th July 2006 UT in queue mode. We used the NIRI f / 6 with a grism with the 4-pix ($\approx$ 0\farcs4 wide) "blue" slit. All 6 sources that were observed were detected in at least one of the $J$ (G0202), $H$ (G0203), and $K$ (G0204) bands. Four sources exhibited continuum emission but no obvious emission lines or breaks. Two sources exhibited strong emission lines and form part of the sample that is presented in this paper. SST24 J143027.1+344007 was targeted in the $J$ (G0202) and $H$ (G0203) bands; SST24 J142800.7+350455 was targeted in the $H$ (G0203) and $K$ (G0204) bands (see Table~\ref{tab:sample}). These grisms correspond to resolutions of R$\approx$650, 940, and 780 for the $J$, $H$, and $K$ filters respectively.  For sky subtraction, the observations were split into 4 minute exposures, and the object was nodded along the slit by 10\arcsec. Typical exposure times were $\approx$1 hour per filter (see Table~\ref{tab:sample}).

\subsection{Data reduction}

The OH telluric sky lines are many orders of magnitude brighter than the signal from our science targets and must be accurately subtracted. We began by removing the first-order background spectrum by subtracting the closest observation in time with a different dither position. Both the NIRSPEC and NIRI low-resolution spectra are tilted relative to the array and are not aligned with the detector columns or rows \citep{mcl98}. We corrected for this by rectifying the spectra. This requires some interpolation between adjacent pixels but makes the fitting of the residual background and extraction of the spectra much simpler. Any residual background sky lines (resulting from the time-variability of telluric sky lines) were then removed by fitting an 15th order polynomial along the lines (masking out the position of the target and any other objects in each frame). The 2-D images were then shifted to move the object spectrum to the same column, and averaged to obtain a final 2-D image. Sigma clipping was used to reject cosmic rays and a rectified bad pixel mask (obtained from the median of multiple dark exposures) was used to reject bad pixels during stacking. The 1-D spectrum was then extracted from the stacked 2-D image and wavelength calibrated using the night sky lines and Argon arcs. All our reductions used standard tasks in the IRAF LONGSLIT package. 

\subsection{Flux calibration}
\label{sec:flux}
Throughout our observing run, we observed telluric standard stars at a similar airmass to our target observations in order to correct for telluric absorption and provide relative flux calibrations of the spectra. We used F0V stars as a compromise between avoiding strong hydrogen lines which are stronger in hotter stars and the lines of other atomic species which become noticeable in cooler stars. For flux calibration, we used the following procedure. We obtained a model spectrum of an F0V star scaled to the magnitude appropriate for the standard star (using the IRAF task PLSPEC). We then divided this by the observed spectrum of the standard star to create a calibration spectrum, which we multiplied into the source spectrum. 

We note that the flux calibration may have large uncertainties due to slit losses. Although the seeing was generally good ($\approx$0\farcs5$-$0\farcs7), the slit was only 0\farcs76 and 0\farcs4 wide for NIRSPEC and NIRI respectively and the targets were too faint to position on the slit precisely. Any offset in the target with respect to the slit will have resulted in an underestimate of the flux density of the source. For 4 sources, we have $K$-band detections from the NDWFS or FLAMEX surveys or from observations by Keck NIRC. In the 2 cases with a significant $K$-band continuum detection, we compare the $K$-band magnitudes with that obtained from the spectra to check our calibrations. For SST24 J142644.3+333051 and SST24 J143424.4+334543, the flux is 0.78 and 0.64 of that expected from their $K$-band magnitudes. For consistency with the observations with no $K-$band magnitudes, we chose not to correct the flux. However, we note that the absolute flux measurements are likely to be underestimated by $\approx$25-50\%.  

\subsection{Emission line fitting}

\begin{figure}[h]
\begin{center}
\includegraphics[height=55mm]{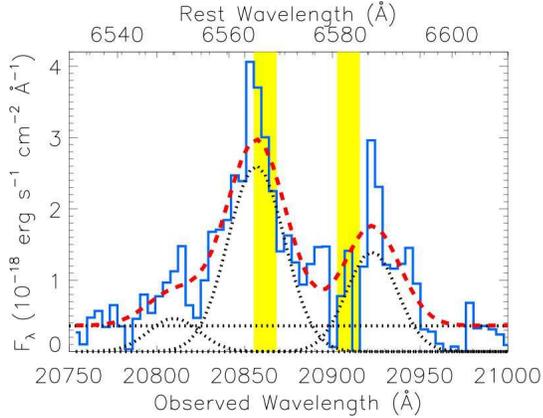}
\end{center}
{\caption[junk]{\label{fig:ex_fit} Example of line fitting for SST24 J143028.5+343221. The blue histogram shows the un-smoothed data. The model fit is shown by the dashed red line. We also show the separate contributions to the model fit from the continuum and individual Gaussian line profiles (black dotted lines). Wavelength regions particularly affected by sky emission lines are shaded in yellow. These regions are given less weight in the fitting process.}}
\end{figure}

We determined line widths and strengths by fitting Gaussian models to the 1-D spectra. These lines are fit simultaneously using {\small MPFITFUN}: a robust non-linear least squares curve fitting routine. Each line is approximated by a Gaussian profile with the width, height and position allowed to vary. The continuum flux is assumed to be constant as a function of wavelength and the height is simultaneously fit with the height, width, and positions of the Gaussian profiles. In order to minimize the number of fitted parameters and help fit the noisier lines, we used the following method. We tied the central wavelengths of the individual lines to have the same redshift. We only allowed the fitted redshift to differ from the redshift measured by eye by 0.03 and constrained the fits to only return positive emission line strengths. We also required all the forbidden lines to have the same widths. We fixed the [NII] $\lambda$6583 / $\lambda$6548 ratio to 3 (as derived by \citealt{sto00}) for the transition probability ratio since these lines are often blended with H$\alpha$ and typically have low signal-to-noise ratios. We also fixed the [OIII] $\lambda$5007 / [OIII] $\lambda$4959 ratio to its theoretical value. We chose not to fit the [SII] $\lambda$$\lambda$ 6717, 6731 lines since, if they are present at all, they have very low signal-to-noise ratios. In cases for which we had measurements of strong [OIII] lines and poor constraints in the [NII] lines, we fixed the width of the [NII] lines to that of the [OIII] lines. The line fitting was weighted by the inverse variance spectrum. This gives less weight to wavelengths which are affected by strong (and often hard to perfectly subtract) sky lines. Figure~\ref{fig:ex_fit} shows an example of how the fitting process works. 

\section{Results}

\subsection{The near-IR spectra}

The final flux-calibrated near-IR spectra are presented in Figure~\ref{fig:kspect}. The fitted line profiles are overplotted. In all 9 spectra covering the appropriate rest wavelengths, H$\alpha$ is prominent and has a highly significant detection over the noise. It is broad ($>$1900 km s$^{-1}$) in 6 cases (we assume that the broad-line component in SST24 J142827.1+354127 dominates the 24$\rm \mu m$ luminosity; see Section~\ref{sec:ind}). [NII] $\lambda\lambda$6548,6583 and [SII] $\lambda\lambda$6717,6731 are also present in many cases, although they are not always significant detections. In the 5 spectra covering the rest wavelengths of H$\beta$ and [OIII] $\lambda\lambda$4959,5007, [OIII] $\lambda$5007 has a significant detection in all but one case, [OIII] $\lambda$4959 has a significant detection in two cases, and H$\beta$ has only a marginal detection in one case. We present the fitted widths and fluxes of the observed emission lines in Tables~\ref{tab:sample}. All quoted line widths and fluxes have been corrected to take into account the instrumental profile of FWHM$_{inst} \approx$10\AA~ and $\approx$14\AA~ for the $H-$ and $K-$band filters respectively on NIRSPEC and FWHM$_{inst} \approx$24\AA,  $\approx$20\AA, and $\approx$28\AA~ for the $J-$, $H-$, and $K-$band filters respectively on NIRI, as measured from the width of the night sky lines. We discuss the individual sources (on order of increasing redshift) in the following section. 

\begin{figure*}[h]
\begin{center}
\includegraphics[height=55mm]{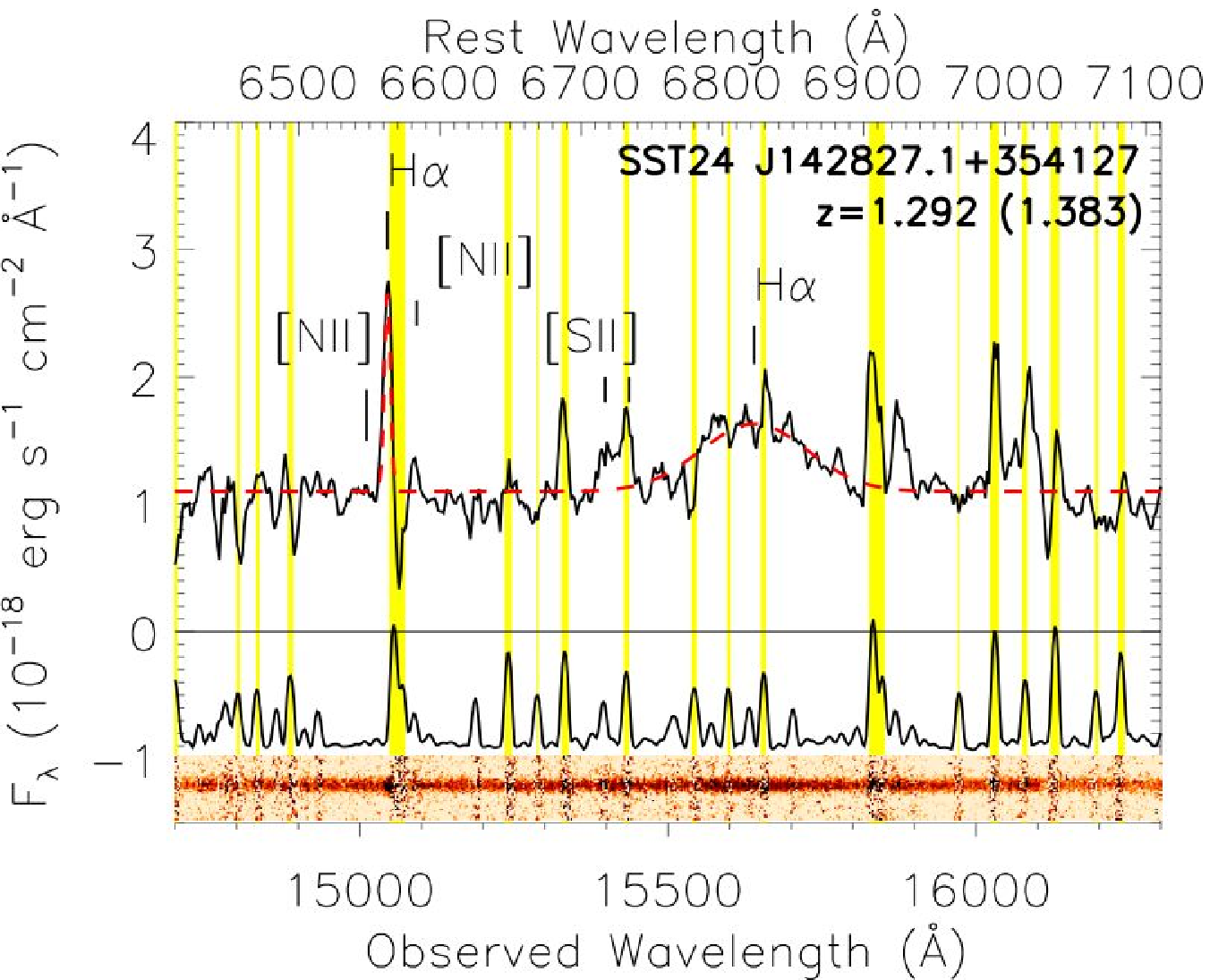}
\includegraphics[height=55mm]{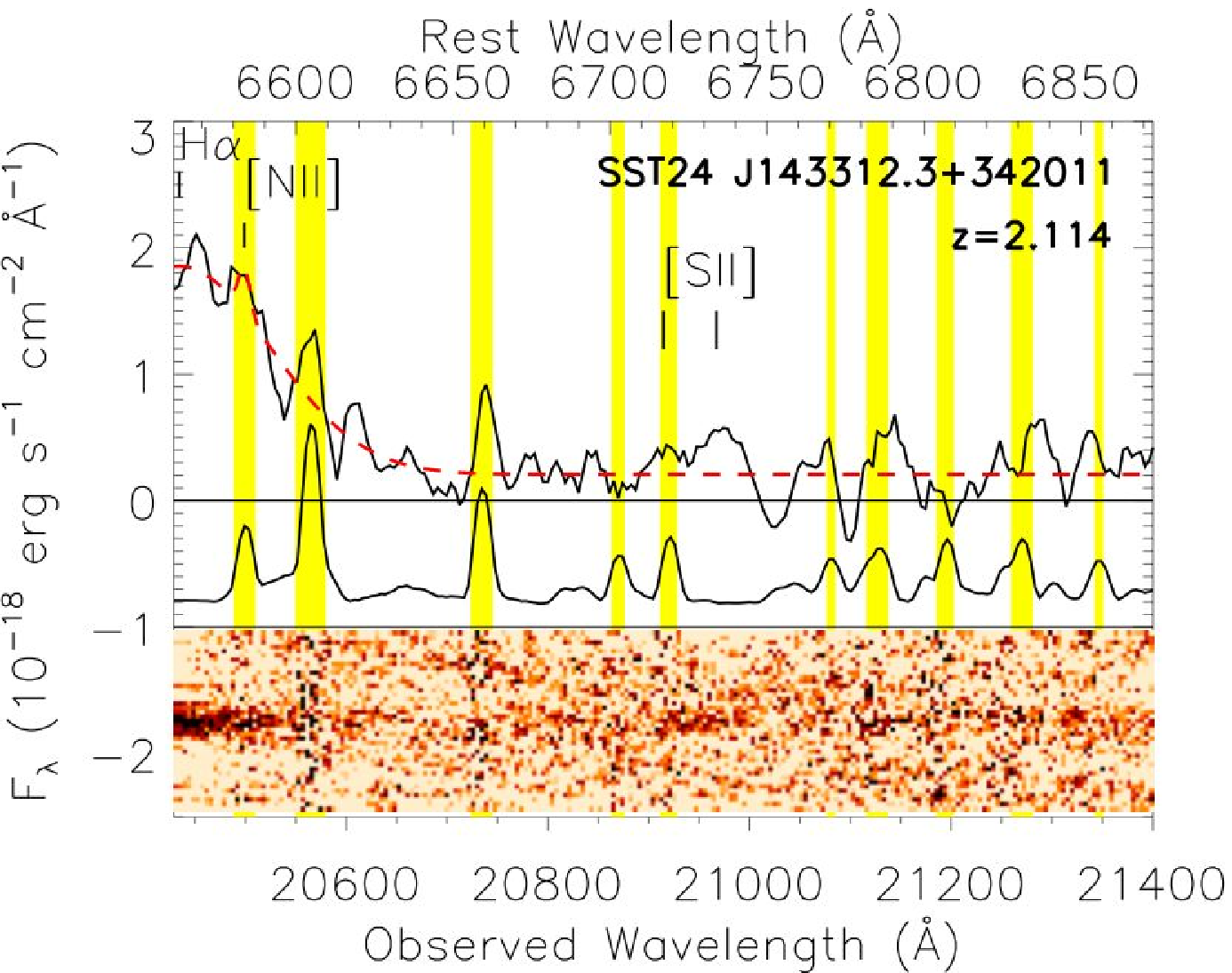}
\includegraphics[height=55mm]{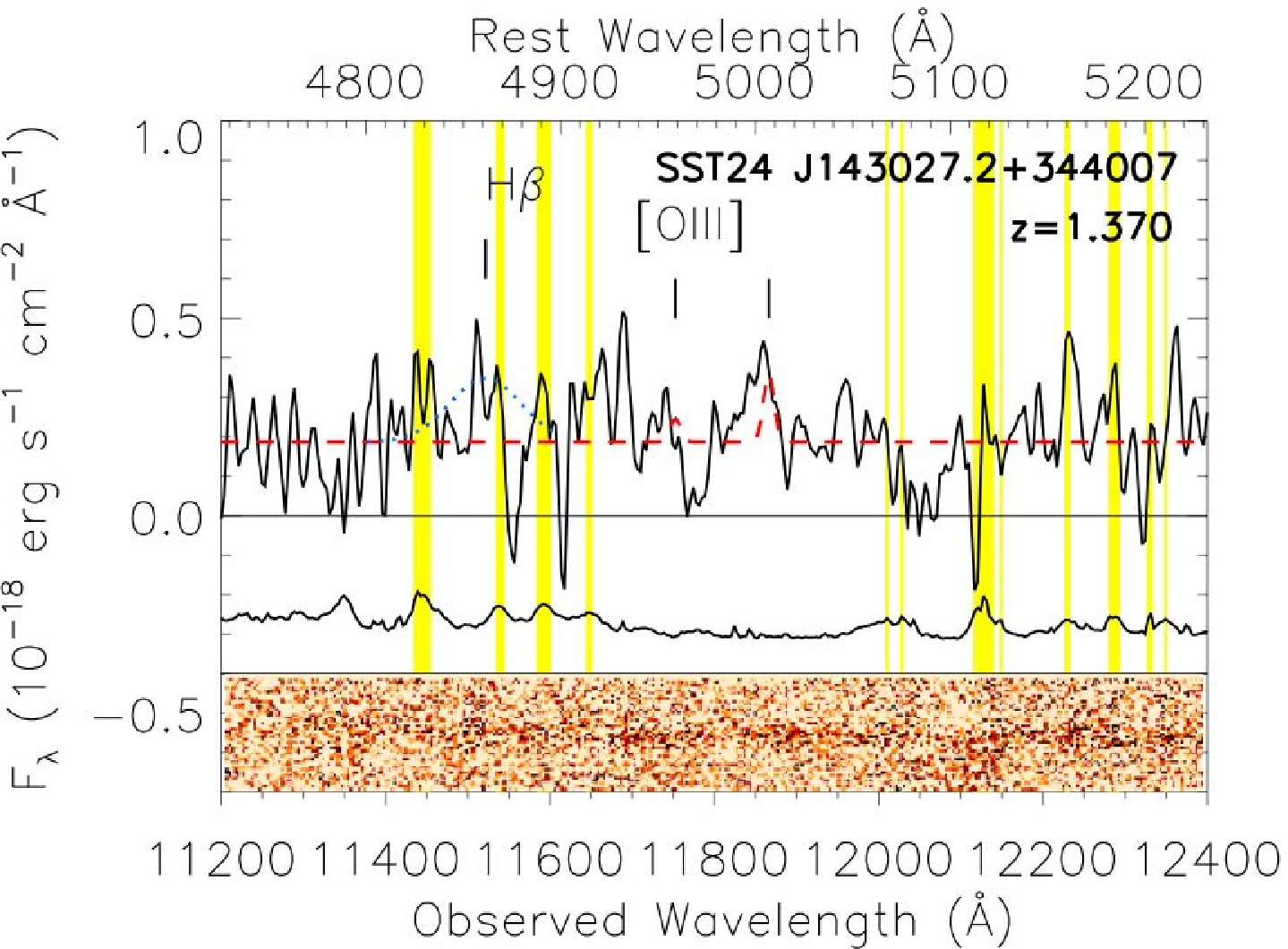}
\includegraphics[height=55mm]{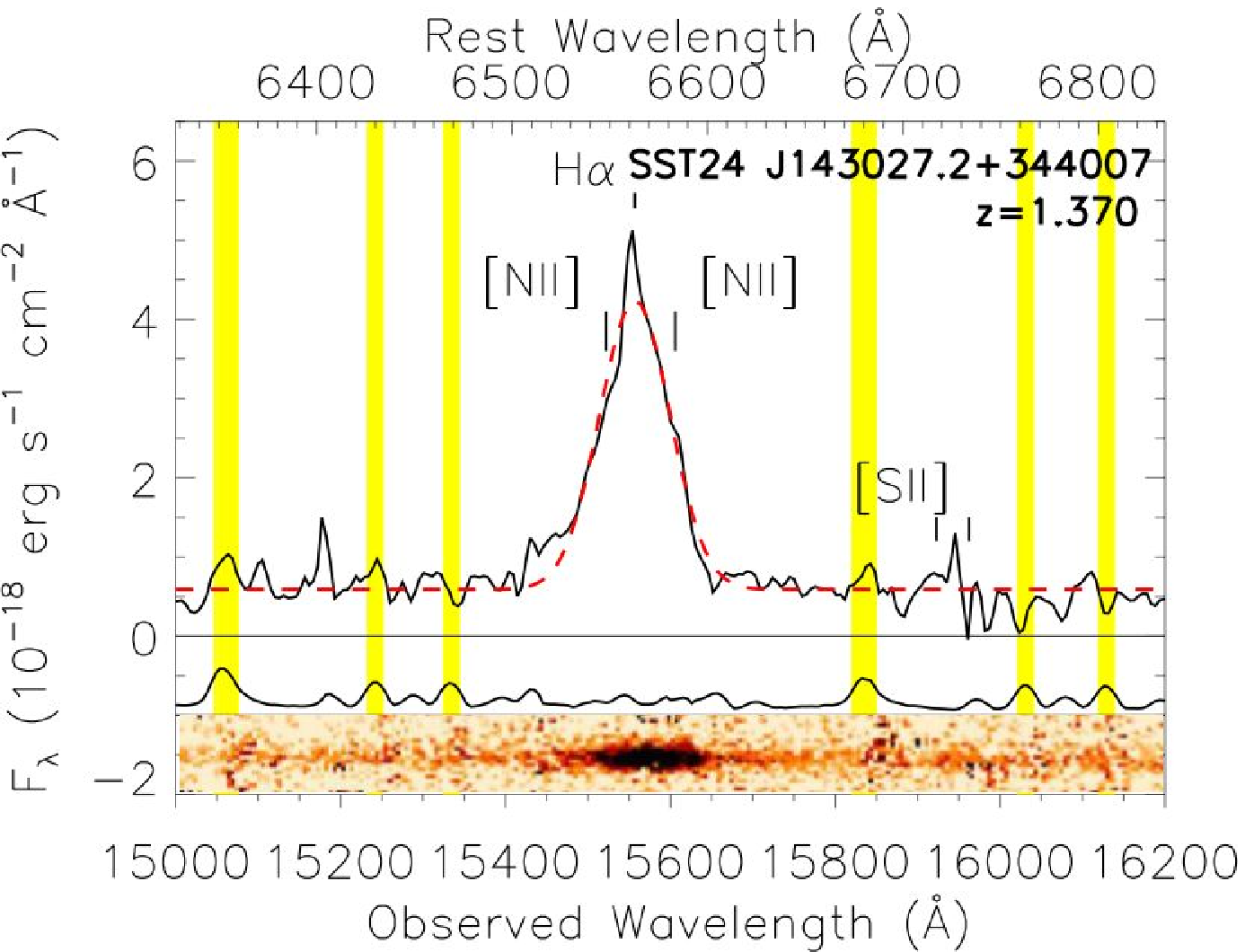}
\includegraphics[height=55mm]{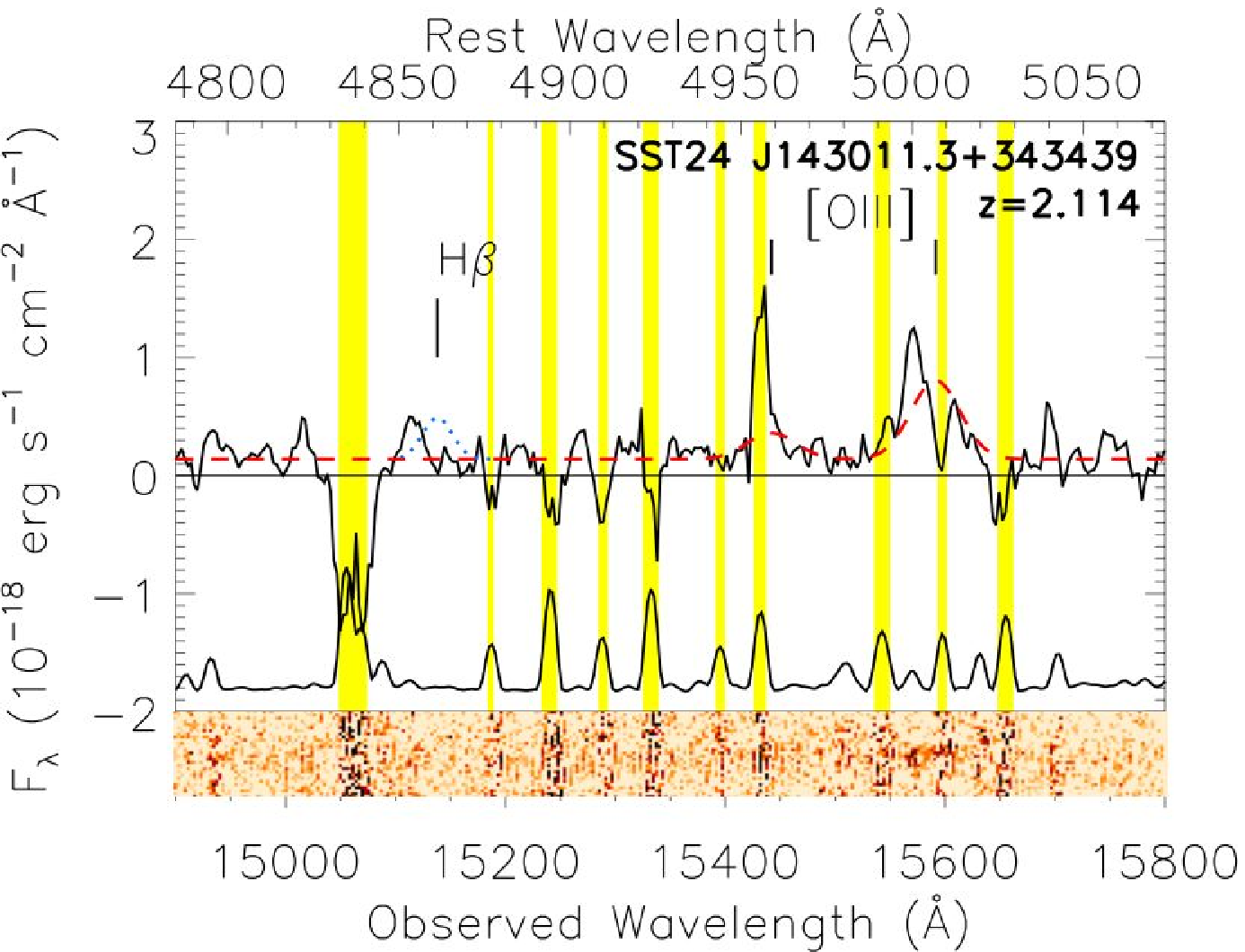}
\includegraphics[height=55mm]{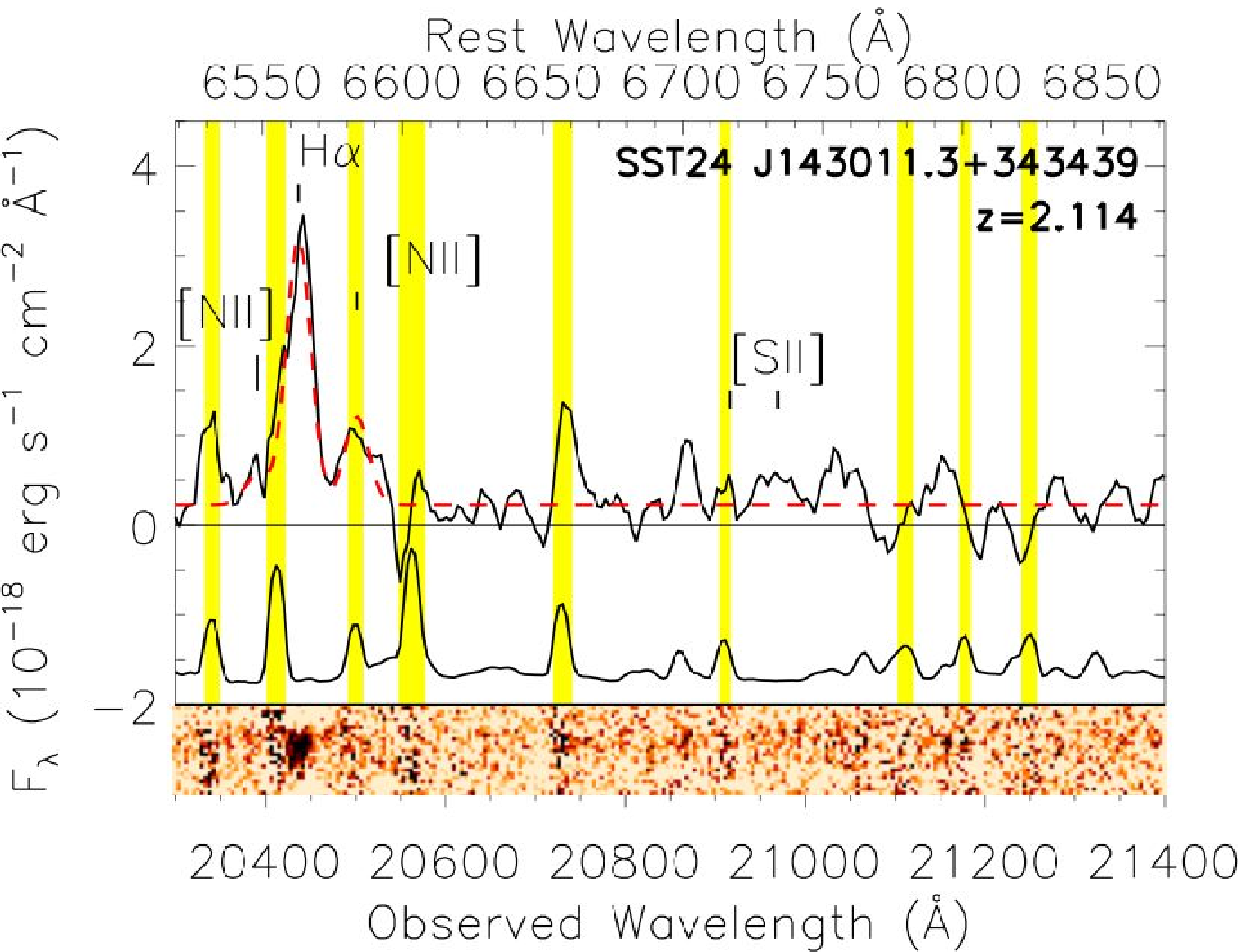}
\end{center}
{\caption[junk]{\label{fig:kspect} Keck NIRSPEC and Gemini NIRI near-IR spectra of the 10 optically faint ULIRGs. The spectra have been smoothed by a 14\AA, 22\AA, 10\AA, 15\AA, and 35\AA~boxcar filter in the NIRSPEC $H-$ and $K-$band filters and NIRI $J$, $H$, and $K$ filters respectively (solid line). The expected positions of typical emission lines are labeled. For each object, the thin solid line (offset from zero by up to 2 $\times 10^{-18} \rm~erg~s^{-1}~cm^{-2}$\AA$^{-1}$ for clarity) shows the corresponding 1 $\sigma$ error spectrum which is dominated by the OH telluric emission lines.  Regions that are particularly affected by the sky emission lines are shaded in yellow. The two-dimensional spectra are also shown with the same wavelength range as that of their 1-D spectra. The fitted line profiles are shown by the dashed line. In the 4 cases in which the H$\beta$ line is not detected, the 3$\sigma$ minimum line profile (with the width fixed to that of H$\alpha$) are shown by the dotted line.}}
\end{figure*}

\addtocounter{figure}{-1}

\begin{figure*}[h]
\begin{center}
\includegraphics[height=55mm]{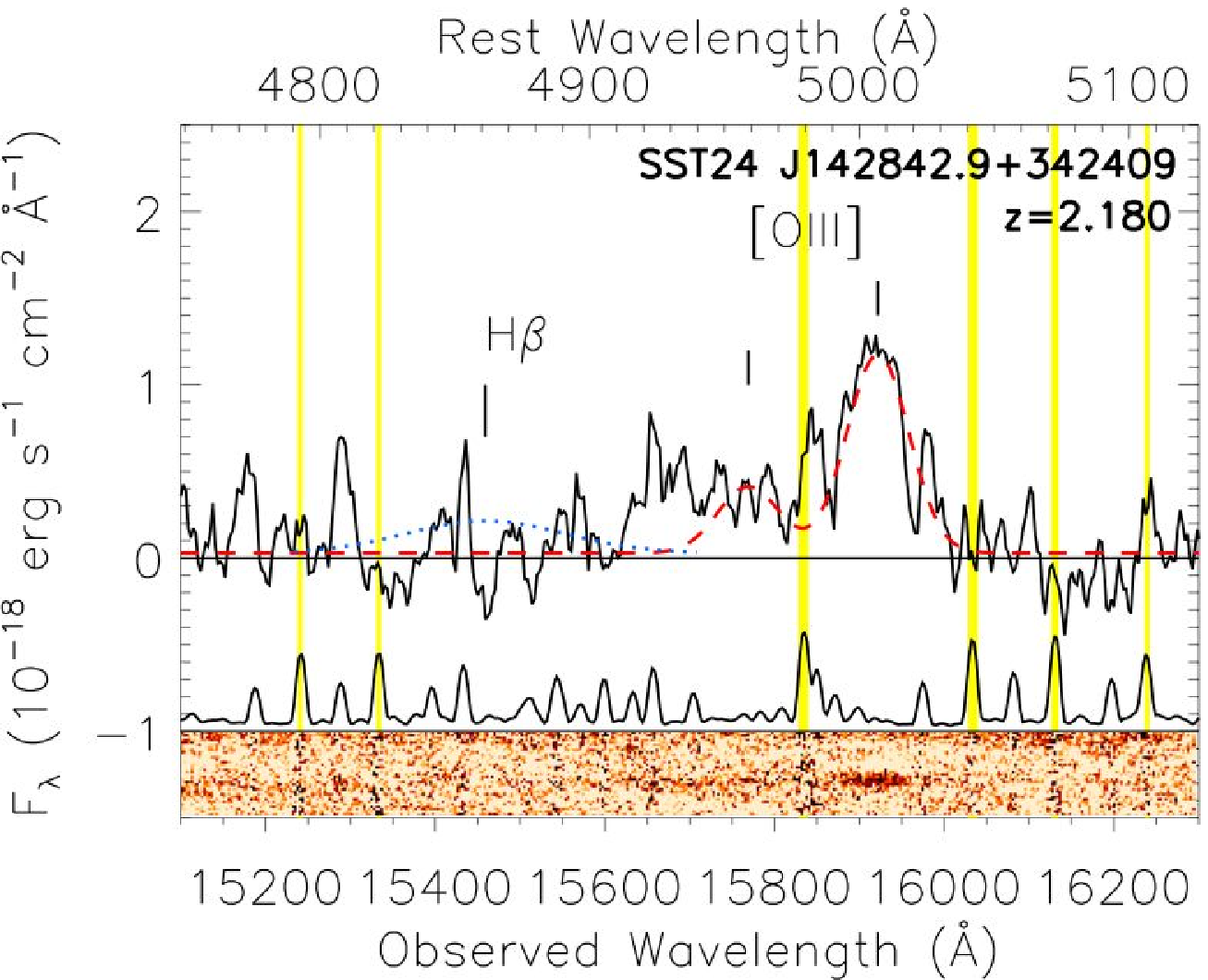}
\includegraphics[height=55mm]{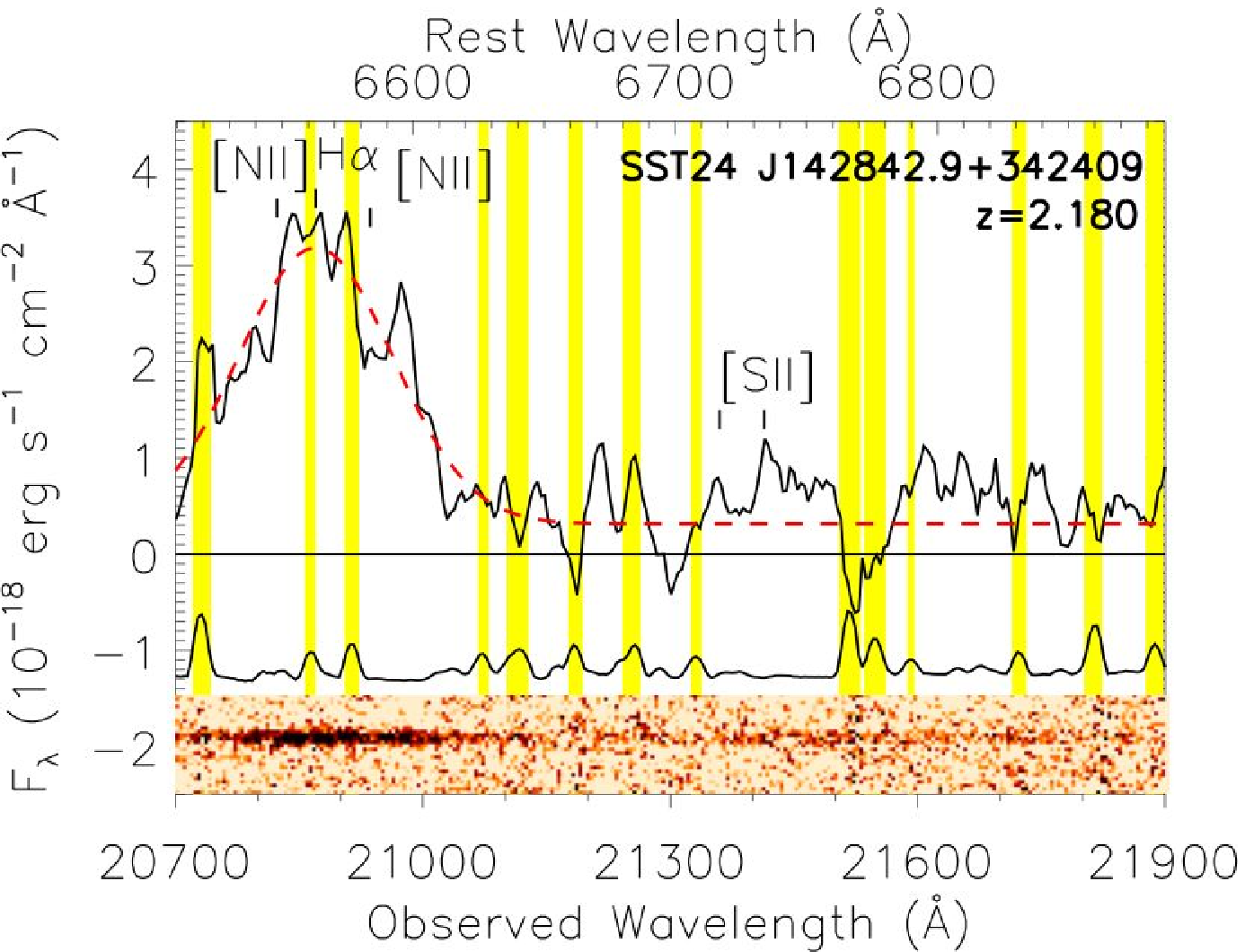}
\includegraphics[height=55mm]{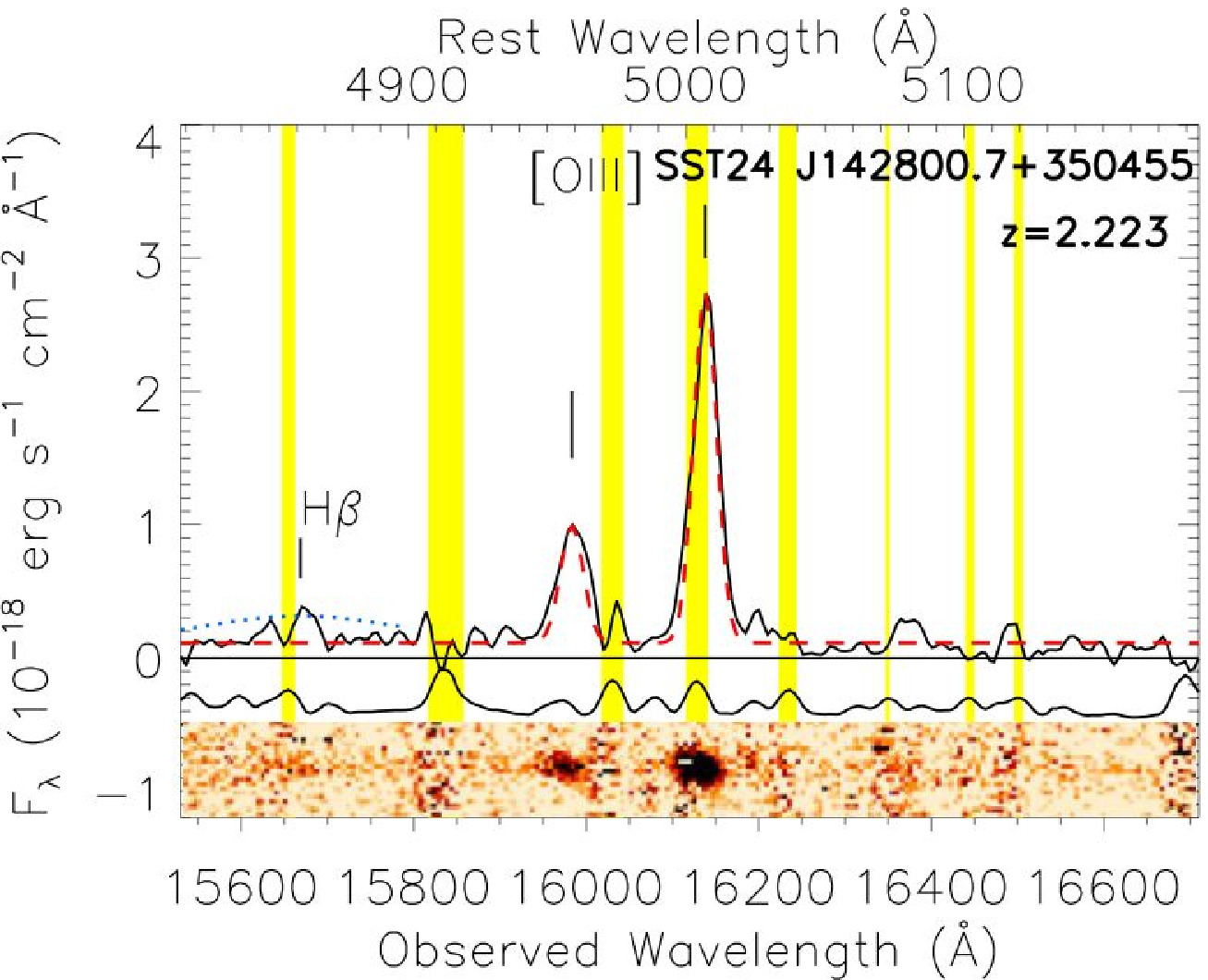}
\includegraphics[height=55mm]{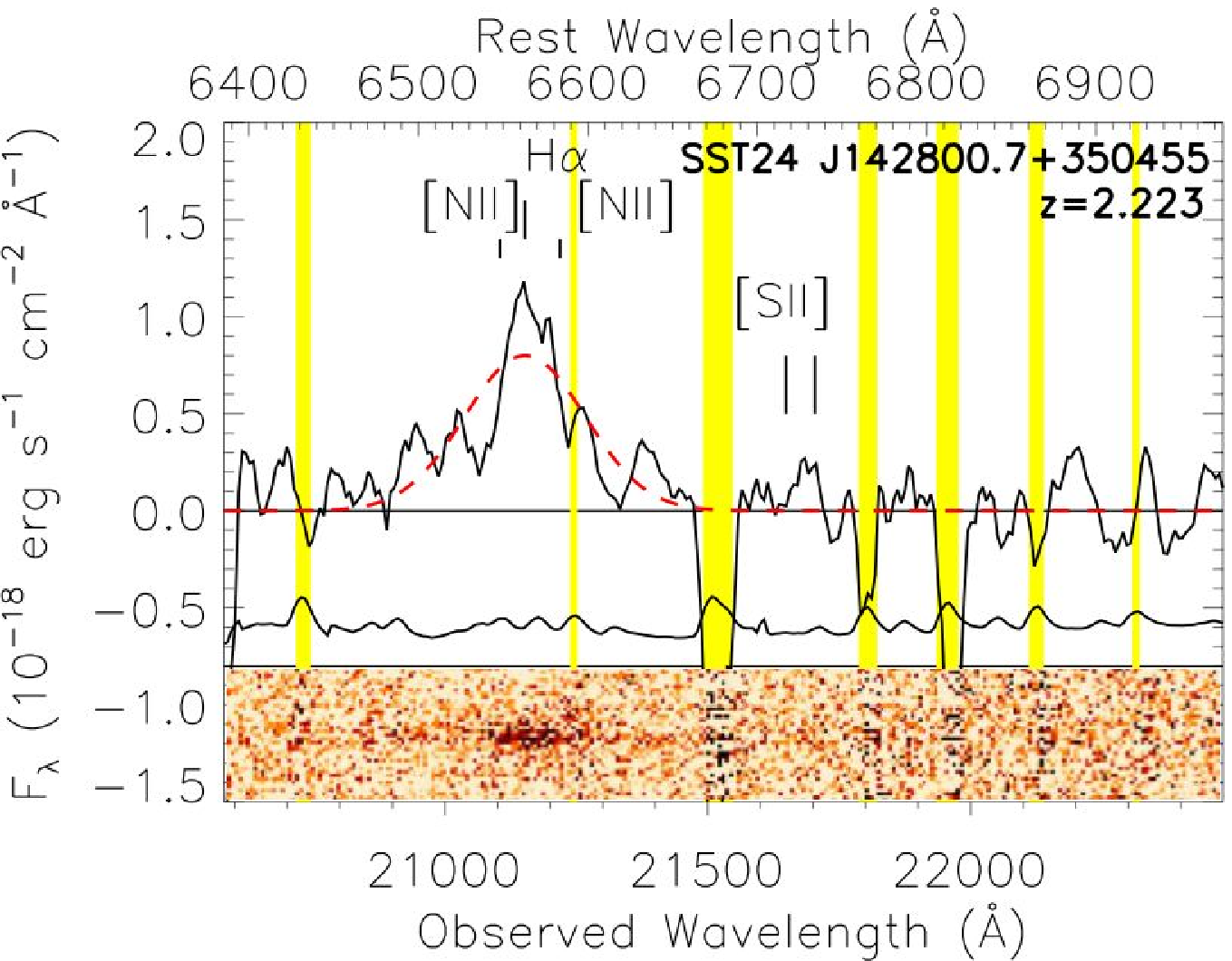}
\includegraphics[height=55mm]{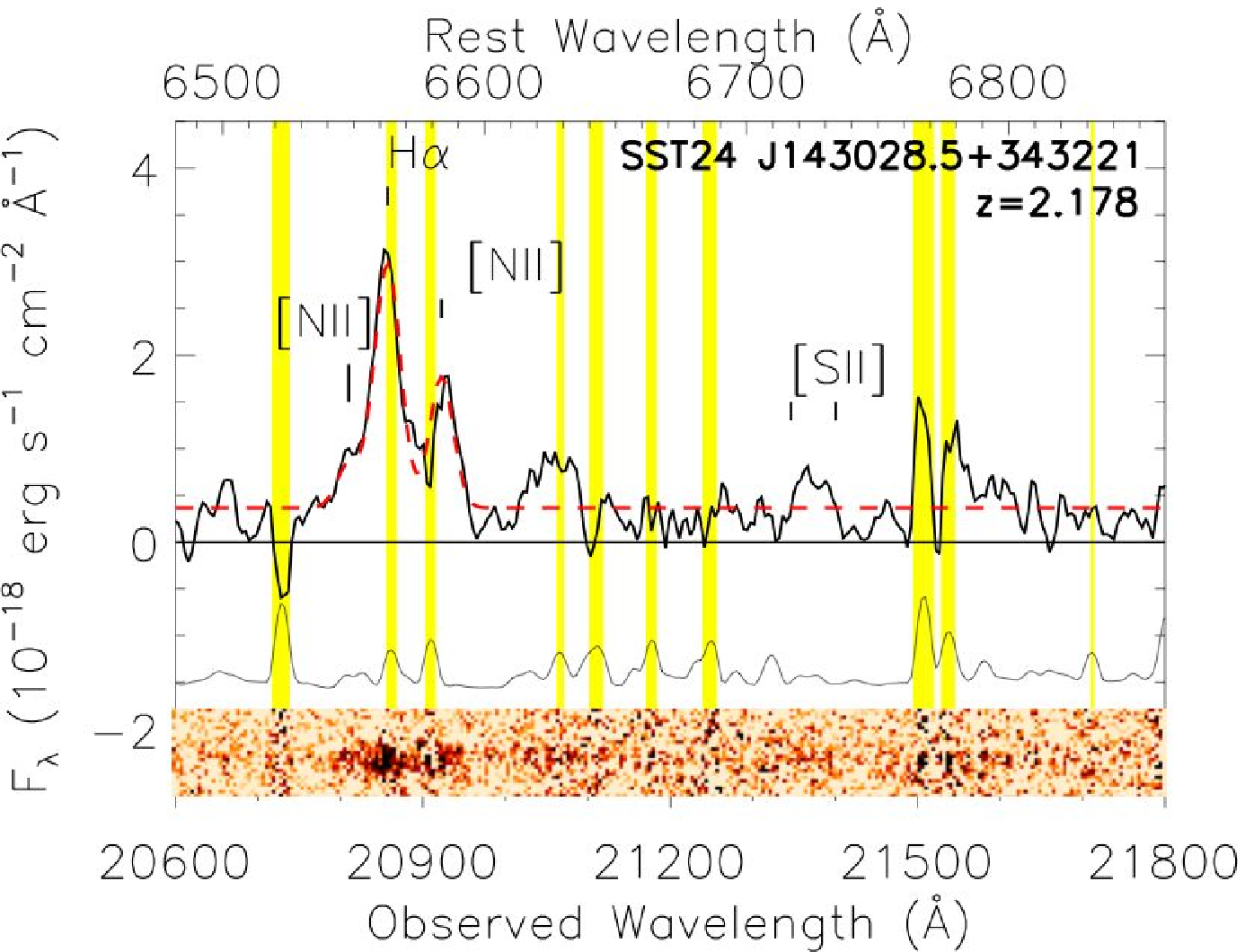}
\includegraphics[height=55mm]{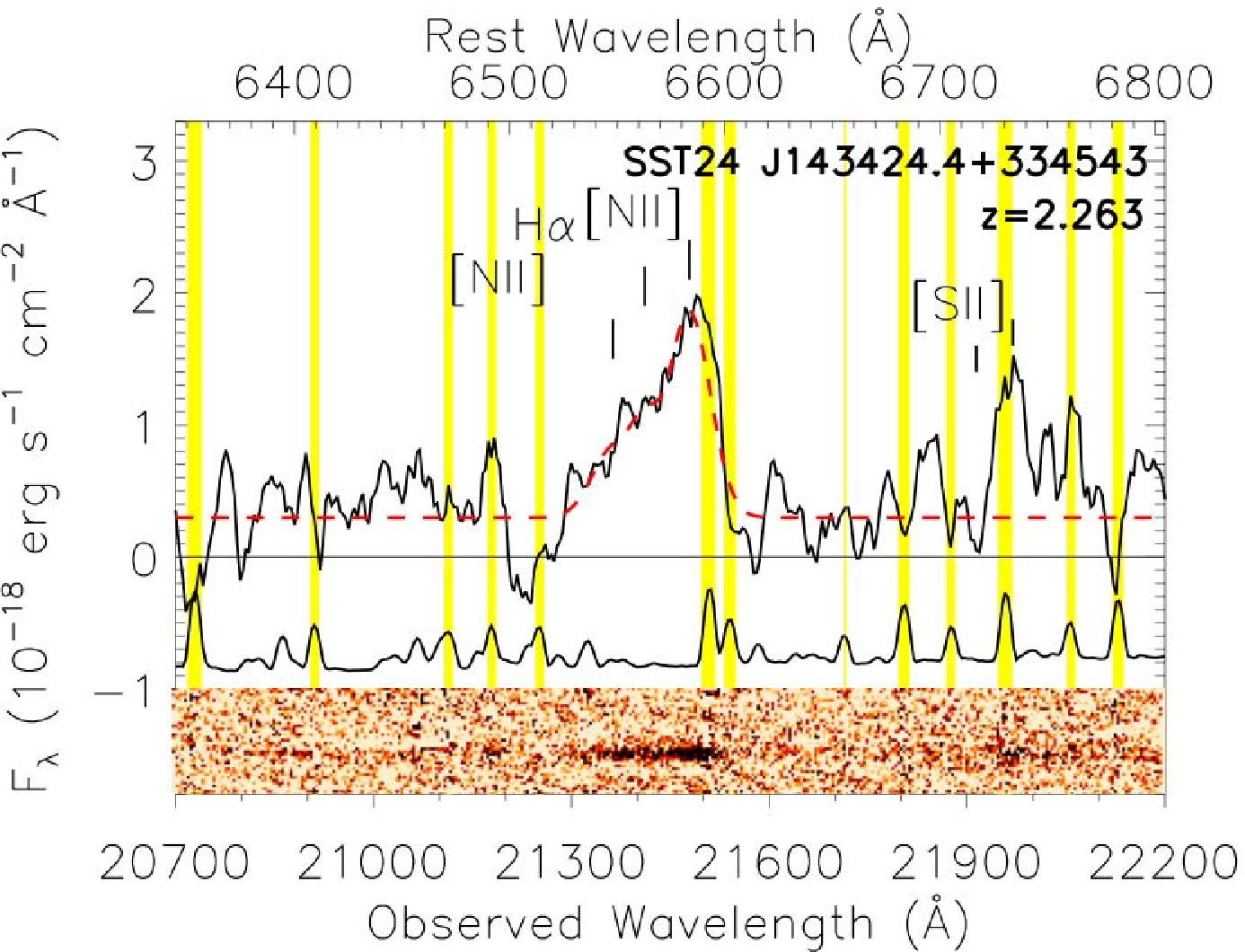}
\includegraphics[height=55mm]{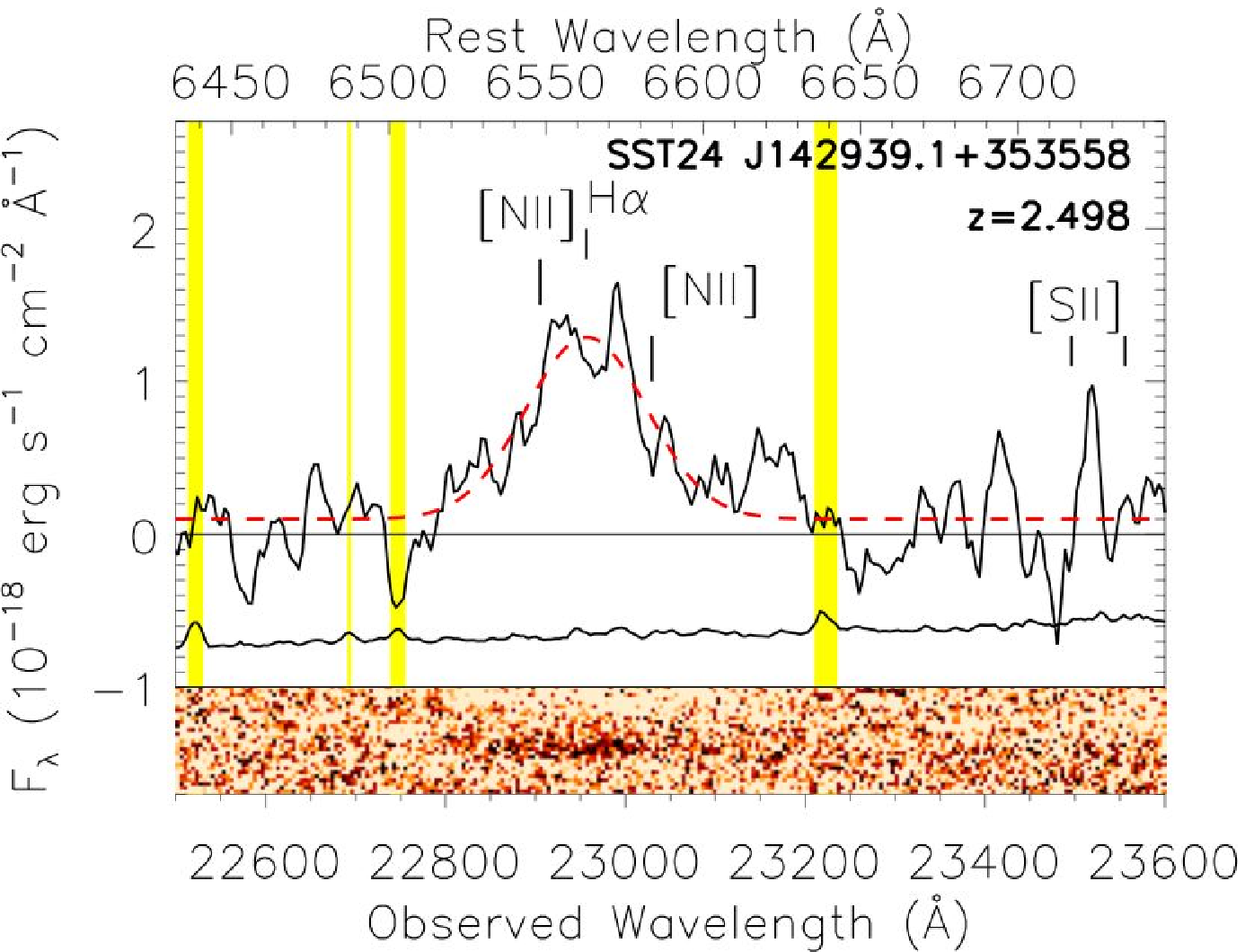}
\includegraphics[height=55mm]{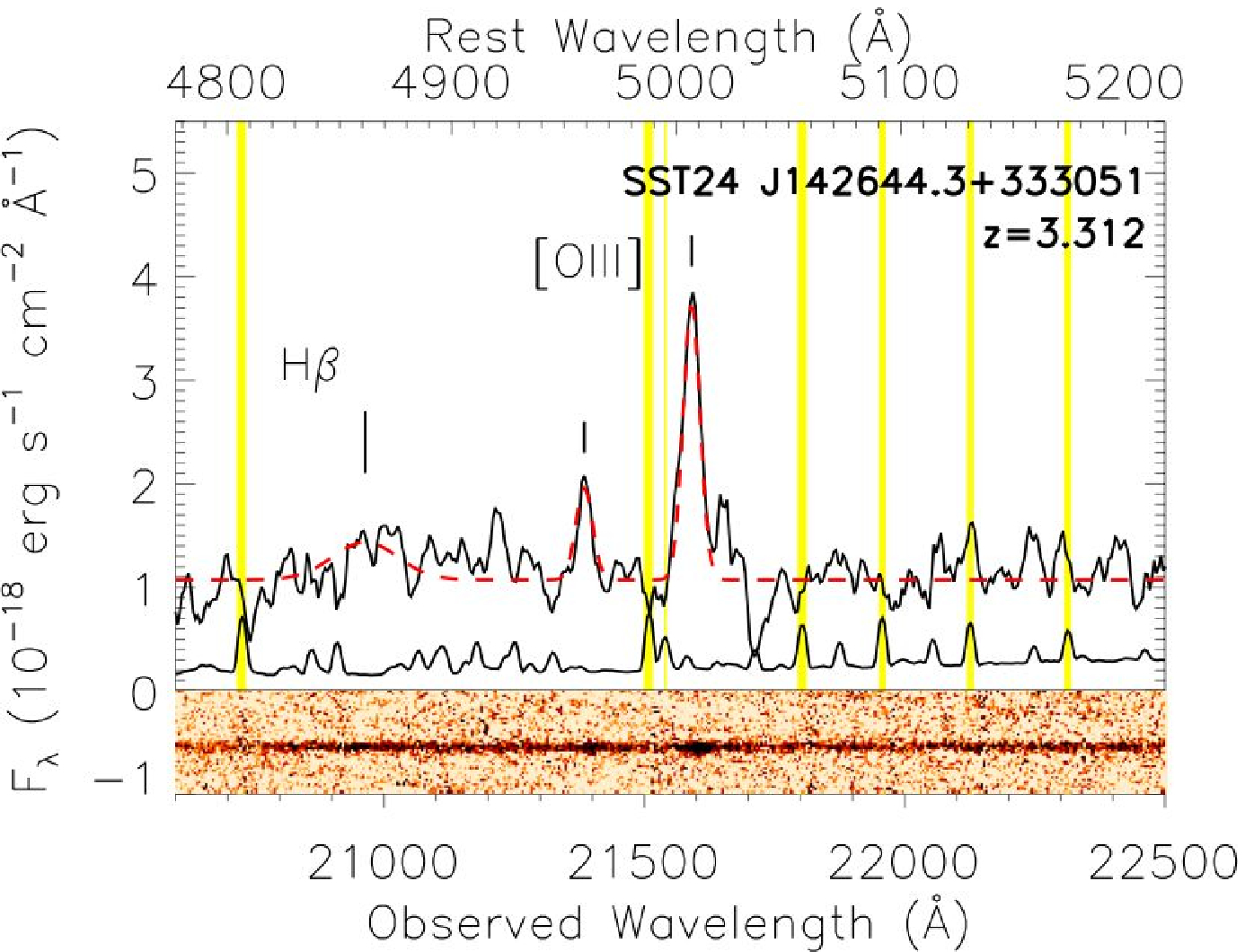}
\end{center}
{\caption[junk]{\label{fig:kspect} continued.}}
\end{figure*}

\subsection{Notes on individual sources}
\label{sec:ind}
\noindent{\it \underline{SST24 J142827.1+354127.}}\textemdash This source is highly unusual in its powerful 24 $\rm \mu m$ emission (f$_{24}$=10 mJy) and faint (but very red) optical magnitude ($B_W-K$=11). The source is discussed in detail in \citet{des06}, where both its IRS and optical spectra are presented. The optical spectrum exhibits a single emission line which is identified as [OII] providing a redshift of $z=$1.292. Given this redshift, SST24 J142827.1+354127 has one of the largest rest-frame 5 $\rm \mu m$ luminosities known ($\rm \nu L_\nu$ = 5.5 $\times 10^{12} L_\odot$), yet its faint optical luminosities and lack of X-ray emission down to the limits of the XBo\"otes survey imply extreme gas and dust obscuration. The IRS spectrum is typical of that of AGN-dominated sources: it exhibits a strong and silicate absorption feature but no PAH features.
Given its known redshift, we observed SST24 J142827.1+354127 in the $H$-band to target the H$\alpha$ line. Our near-IR spectrum shows significant continuum and interestingly, reveals two redshift components which we denote M142827a and M142827b. 
The redshift of M142827a agrees with that of the optical spectrum. Unfortunately, a powerful night sky line happens to fall at the same wavelength as the observed wavelengths of the diagnostically important lines, H$\alpha$ and [NII] $\lambda6583$, making estimates of the H$\alpha$ flux and width very uncertain. Given the effect of the sky line, our determination of FWHM$_{H\alpha} \approx$230 $\rm km~s^{-1}$ is highly uncertain. However, it is unlikely to be broad enough to constitute a broad AGN emission line. The [NII]/H$\alpha$ ratio is poorly constrained and so we classify this component as dominated by either a starburst or type 2 AGN.
M142827b is newly discovered in our NIRSPEC observations. It is revealed by a broad line at an observed wavelength of 15,689 \AA. If this is H$\alpha$, then the source is at $z=$1.383. Given the width of the line, the source must be dominated by a type 1 AGN. Although there is no evidence for a second source in the optical spectrum, the IRS redshift estimated from the silicate absorption feature falls between the two NIRSPEC redshifts as one would expect for a composite source in which both components were optically faint. 
Given the large velocity separation between the two redshift components ($\Delta$v$\approx$11,500 km s$^{-1}$), we conclude that the mid-IR flux is coming from two separate sources which happen to be coincident along our line of sight. If SST24 J142827.1+354127 is a composite source, this would certainly explain its unusually high infrared flux. Because the H$\alpha$ luminosity of the broad line is $\approx$10 times larger than that of the narrow H$\alpha$ line of the lower redshift source, we assume that M142827b is the dominant source of the infrared flux, and hence the source which satisfies our $R-[24]>14$ criterion. \\

{\it \underline{SST24 J143027.1+344007.}}\textemdash This source was observed with Gemini  NIRI because of its bright $K$-band magnitude ($K$=17.89). It exhibits broad H$\alpha$ emission (FWHM$_{H\alpha}$=1940 $\rm km~s^{-1}$) in the $H$-band confirming that it is dominated by an AGN, and providing a redshift of $z$=1.370. Our observations in the $J$-band show a marginal detection of [OIII]$\lambda$5007 and no H$\beta$ detection. The source exhibits significant continuum emission and so cannot have simply been missed off the slit. The weakness of [OIII] emission is unusual in comparison to the other sources in our sample. Using the minimum H$\alpha$ to H$\beta$ ratio, we estimate at least 4.6 magnitudes of extinction at the wavelength of H$\alpha$. The unattenuated H$\alpha$ luminosity must be at least 3$ \rm \times 10^{44}~ergs~s^{-1}$. \\

{\it \underline{SST24 J143312.7+342011.}}\textemdash This source was observed in the $K$-band by Keck  NIRSPEC because its redshift ($z=2.2$; as estimated from the silicate absorption feature) put the observed wavelength of H$\alpha$ in the $K$-band. Apart from a moderate silicate absorption feature, the IRS spectrum is featureless. It is best fit by an AGN-dominated template. Our NIRSPEC observations show that the source has a slightly lower redshift of $z=$2.114, putting $H\alpha$ at the very edge of our wavelength coverage. We do however, appear to have observed the peak of the line, so our estimates of the line strength and width should not be too severely affected. The H$\alpha$ line is very broad (FWHM$_{H\alpha}$=3150$\pm$300 km s$^{-1}$) with a large equivalent width confirming that the source is dominated by an AGN. It has the second brightest H$\alpha$ luminosity of our sample. The presence of [NII] is not significant given the large uncertainties introduced by a coincident night sky line. The source appears to have [SII] line emission, although at fairly low significance. \\

{\it \underline{SST24 J143011.3+343439.}}\textemdash This source was chosen because its redshift from Keck optical spectroscopy ($z$=2.12) put the observed wavelength of H$\alpha$ in the $K$-band. It was observed in both the $K-$ and $H-$bands. With $R-[24]=12.5$, this source does not strictly obey our $R-[24]>14$ selection criteria and it has not been observed with IRS. The near-IR spectrum confirms the optical redshift ($z$=2.114). It exhibits relatively narrow H$\alpha$ (FWHM=470 km s$^{-1}$) and its combined [NII]/H$\alpha$ ratio and [OIII]/H$\beta$ limits suggests that it is a type-2 AGN. From the minimum H$\alpha$ to H$\beta$ flux ratio, we estimate at least 2.4 magnitudes of extinction at the wavelength of H$\alpha$. \\

{\it \underline{SST24 J143028.5+343221.}}\textemdash This source was originally chosen for near-IR spectroscopy because optical follow-up on Keck LRIS found a redshift of $z=2.178$ (from Ly$\alpha$). It was then subsequently observed with $Spitzer$ IRS which confirmed this redshift. The source exhibits relatively narrow H$\alpha$ (FWHM=542 km s$^{-1}$) and its large [NII]/H$\alpha$ ratio suggests it is a type-2 AGN. The IRS spectrum shows a deep silicate absorption feature and possible PAH emission features (see Hidgon et al. in preparation for more details). \\

{\it \underline{SST24 J142842.9+342409.}}\textemdash This source was chosen because it exhibited a featureless power-law spectrum in the mid-IR. Follow-up optical spectroscopy also failed to yield a redshift. More sophisticated fitting of the IRS spectrum in \citet{wee06} estimated a possible redshift of $z=$1.1. We observed it in the $K$-band and detected a strong H$\alpha$ line, identifying the source as AGN-dominated and putting it at a redshift of $z$=2.180. We then followed up in the $H$-band to target the [OIII] and H$\beta$ lines. The H$\alpha$ line is broad ($\approx$3000 km s$^{-1}$) with a large equivalent width and the observed H$\alpha$ luminosity of $\approx 2 \times 10^{43} \rm ergs~s^{-1}$ is the largest in the sample.  The [OIII] line is unusually broad (FWHM$_{[OIII]}$=1600 $\rm km~s^{-1}$), perhaps due to shock-induced superwind activity (cf. \citealt{tak06}). The $H-$band spectrum shows no significant H$\beta$ flux suggesting large columns of attenuating dust (see Section~\ref{sec:dust}). Given the new redshift, we re-examine the IRS spectrum but find no evidence for any silicate absorption in the source. This is somewhat surprising given the large columns of dust which must exist in this source. See Section~\ref{sec:midir} for further discussion of this source. \\

{\it \underline{SST24 J142800.7+350455.}}\textemdash This source was observed with Gemini NIRI because of its bright $K$-band magnitude ($K$=17.78).  It has not been observed with IRS. The strong [OIII] and H$\alpha$ lines detected in the $H-$ and $K-$bands respectively determine that $z$=2.223. Although the best obtained fit gives FWHM$_{H\alpha}$ = 3833 $\rm km~s^{-1}$, other (slightly worse) fits indicate a narrower line profile. Nevertheless, all fits suggest FWHM$_{H\alpha}>$2000 $\rm km~s^{-1}$ and the source is classified as a type I AGN. Because the H$\alpha$ flux is relatively low and FWHM$_{H\alpha}$ is relatively high, we can only put weak constraints on the extinction. \\

{\it \underline{SST24 J143424.4+334543.}}\textemdash This source was chosen because, although the IRS spectrum showed a featureless power-law spectrum, optical follow-up spectroscopy found a redshift of $z=2.26$. The NIRSPEC spectrum confirms the optical redshift and shows a very large [NII]/H$\alpha$ ratio. Although difficult to measure given the powerful [NII] emission, the width of the H$\alpha$ line is relatively narrow and we classify it as a type 2 AGN. The IRS spectrum shows a downward trend in the flux at $\approx$31$\rm \mu m$ where one would expect the silicate absorption feature to lie. It may be that the feature is too near the edge of the spectrum to yield a convincing IRS redshift. The IRS spectrum shows possible PAH emission features although it is fairly noisy. It is interesting that the only sources with IRS spectra showing possible PAH emission features are the 2 sources which are optically classified as type 2 AGN. \\

{\it \underline{SST24 J142939.1+353558.}}\textemdash This source was observed with IRS and Keck LRIS but no redshift was determined. We obtain a redshift of $z$=2.498. The H$\alpha$ line is broad and we classify the source as a type 1 AGN. The source is also detected in the Chandra XBo\"otes survey (\citealt{ken05}; \citealt{mur05}; \citealt{bra06b}) with an X-ray luminosity of L$_{0.5-7keV}$=7.7$\times 10^{42}~\rm ergs~s^{-1}$. The IRS spectrum did not yield a redshift because the silicate absorption feature falls at the very edge of the spectrum.\\

{\it \underline{SST24 J142644.3+333051.}}\textemdash We targeted this source because of its featureless IRS spectrum. We observed it in the $K$-band to target the H$\beta$ and [OIII] lines, given its optical redshift of $z=3.355$ obtained on Keck LRIS (see Desai et al. in preparation). We obtain a slightly lower redshift of $z=3.312$ from the [OIII] lines. Although only marginally detected, the H$\beta$ line is broad (FWHM =2066$\pm$394 km s$^{-1}$) and we classify this source as a type 1 AGN. SST24 J142644.3+333051 is also a powerful X-ray source with L$_{0.5-7keV}$=9$\times 10^{44} \rm ergs~s^{-1}$, confirming that it must be a powerful AGN. The silicate absorption feature falls beyond the observed wavelength range of the IRS spectrum.\\

\section{The characteristics of optically faint ULIRGs}

In this section, we use line diagnostics to characterize the sources into broad-line type 1 AGN, narrow-line type 2 AGN or starburst-dominated sources. We derive physical properties and in cases with both H$\alpha$ line measurements and upper limits for the H$\beta$ line flux, we estimate lower limits to the dust extinction. We also discuss the mid-IR spectra.

\subsection{Line diagnostics}
\label{sec:linediag}
Six sources show broad ($>$1900 km s$^{-1}$) H$\alpha$ lines which are characteristic of AGN-dominated sources. For the one source observed at shorter rest-frame wavelengths (SST24 J142644.3+333051), the H$\beta$ line is only marginally detected but also appears to be broad. We also classify this source as a type 1 AGN. A common diagnostic to distinguish between AGN- and starburst-dominated sources uses the line ratios of [NII]/H$\alpha$ and [OIII]/H$\beta$ as indicators of the hardness of the EUV radiation field in the narrow-line regions (\citealt{bal81}; \citealt{ost85}; \citealt{vei87}; \citealt{kau03}; \citealt{kew06}). Unfortunately, these diagnostics are unreliable for sources with broad Balmer lines since the broad-line component, which likely arises in the broad-line region (BLR), typically overwhelms any narrow-line component. Three of our sources show narrow-line H$\alpha$ emission and have significant [NII] emission, enabling reliable measurements of the ratio. The [NII]/H$\alpha$ ratios are 0.33, 0.48, and 4.6. AGN-dominated sources typically have [NII]/H$\alpha >$0.7 (e.g., \citealt{swi04}) suggesting 1 of our sources is a type 2 AGN. The source with [NII]/H$\alpha$=0.33 has a high [OIII]/H$\beta$ ratio. The [NII]/H$\alpha$ vs. [OIII]/H$\beta$ classification schemes of \citet{kew01} and \citet{kau03} suggest that this source is also a type 2 AGN. Our limited statistics therefore suggest that both the broad and narrow-line sources in our sample are AGN-dominated.

\subsection{Physical properties}
\label{sec:phys}

The main derived physical properties of the sources are shown in Table~\ref{tab:sample}. The sources have H$\alpha$ luminosities in the range $1 \times 10^{42} - 2.4 \times 10^{43} \rm ergs ~s^{-1}$.  For sources classified as type 1 AGN, the SMBH mass can be estimated from both the FWHM and luminosity of the H$\alpha$ line using the prescription of \citet{gre05}. We estimate values of 0.2$-$1.3 $\rm \times~10^{8} M_{\odot}$ which are surprisingly low for such bolometrically luminous sources. However, as we discuss in Section~\ref{sec:dust}, these values are likely to be severely underestimated due to attenuation by dust. To determine the infrared luminosities from the observed 24 $\rm \mu m$ luminosities, we assume a bolometric correction of 7.0 (typical of AGN-dominated sources; \citealt{xu01}). All our sources have $\rm L_{IR} > 10^{12} L_\odot$ and are therefore classified as ULIRGs. Two sources are detected in the Chandra XBo\"otes survey (SST24 J142939.1+353558 and SST24 J142644.3+333051 with X-ray luminosities of L$_{0.5-7keV}$=7.7$\times 10^{42}~\rm ergs~s^{-1}$ and 9$\times 10^{44} \rm ergs~s^{-1}$ respectively. 
For SST24 J142939.1+353558, assuming a canonical accretion efficiency of $\epsilon$=0.1 and that the correction from the X-ray luminosity to the bolometric luminosity is a factor of 35, the accretion rate onto the SMBH is $\approx 5 \times 10^7 \rm M_{\odot} Gyr^{-1}$ \citep{bar01}. Assuming the H$\alpha$ line is attenuated by a factor of 10 in flux, its black hole mass is $\approx 10^8 M_{\odot}$. If the source has been accreting at a similar rate throughout its lifetime, it would have taken it $\approx$ 2 Gyrs to build up to this mass. Given this large value, we suggest that it is likely that the X-ray luminosity is also attenuated by large columns of gas associated with the dust. This is also supported by the fact that only 2/10 of the sources are detected in the XBo\"otes survey even though they are classified as powerful AGN. 

\subsection{Dust extinction}
\label{sec:dust}

Observations of the Balmer decrement can provide an estimate of the extinction due to dust in these sources and hence the correction one must apply to obtain intrinsic line luminosities. \citet{gas84} show theoretically that for AGN, H$\alpha$/H$\beta \approx$3.1. This is slightly higher than the expected value of H$\alpha$/H$\beta$=2.86 for case B recombination at T=10,000K and $n_e\approx$10$^4$ cm$^{-3}$ \citep{ost89}. We assume H$\alpha$/H$\beta$=3.1 for the intrinsic (unreddened) ratio. In three cases (SST24 J143027.1+344007, SST24 J143011.3+343439, and SST24 J142842.9+342409), we have significant detections of the H$\alpha$ line flux and useful upper limits for the H$\beta$ line flux. We measure 3$\sigma$ upper limits on the H$\beta$ flux from the noise spectrum assuming the same FWHM as measured for the H$\alpha$ line (see Figure~\ref{fig:kspect}). The upper limits are given in Table~\ref{tab:sample} and correspond to H$\alpha$/H$\beta >$ 22.5, 8.5, and 15.7 for SST24 J143027.1+344007, SST24 J143011.3+343439, and SST24 J142842.9+342409 respectively. This demonstrates that both the broad- and narrow-line sources suffer from significant optical extinction. Assuming the standard extinction curve of \citet{ost89}, we obtain E(B-V)$>$ 1.9, 1.0, and 1.6, corresponding to extinctions of A(H$_\alpha$)$>$4.6, 2.4, and 3.8 magnitudes. This corresponds to a correction factor of at least , 69, 9, and 33 times the H$\alpha$ luminosities quoted in Table~\ref{tab:sample}.  The extinction limits are far larger than that inferred for composite spectra of reddened quasars from SDSS \citep{ric03} and larger than even the largest extinctions inferred for UV-selected galaxies (A(H$_\alpha$)=1.7; \citealt{erb06}) and distant red galaxies (A(H$_\alpha$)=1.7; \citealt{van04}) at similar redshifts. If the true extinctions are close to the measured limits, although they sample sizes are small, they may be comparable to sub-millimeter galaxies \citep{tak06},  local dusty starbursts (A(H$_\alpha$)=2.9; \citealt{pog00}, and some of the largest optical extinctions of the most extreme local ULIRGs by \citet{vei95}. Assuming that all the sources suffer a similar extinction, the line widths and corrected H$\alpha$ luminosities of $\gtsimeq 10^{43}-10^{44} \rm ~ergs~s^{-1}$ suggest SMBH masses of $\gtsimeq 10^8-10^9 \rm M_\odot$. 

\begin{figure}[h]
\begin{center}
\includegraphics[height=50mm]{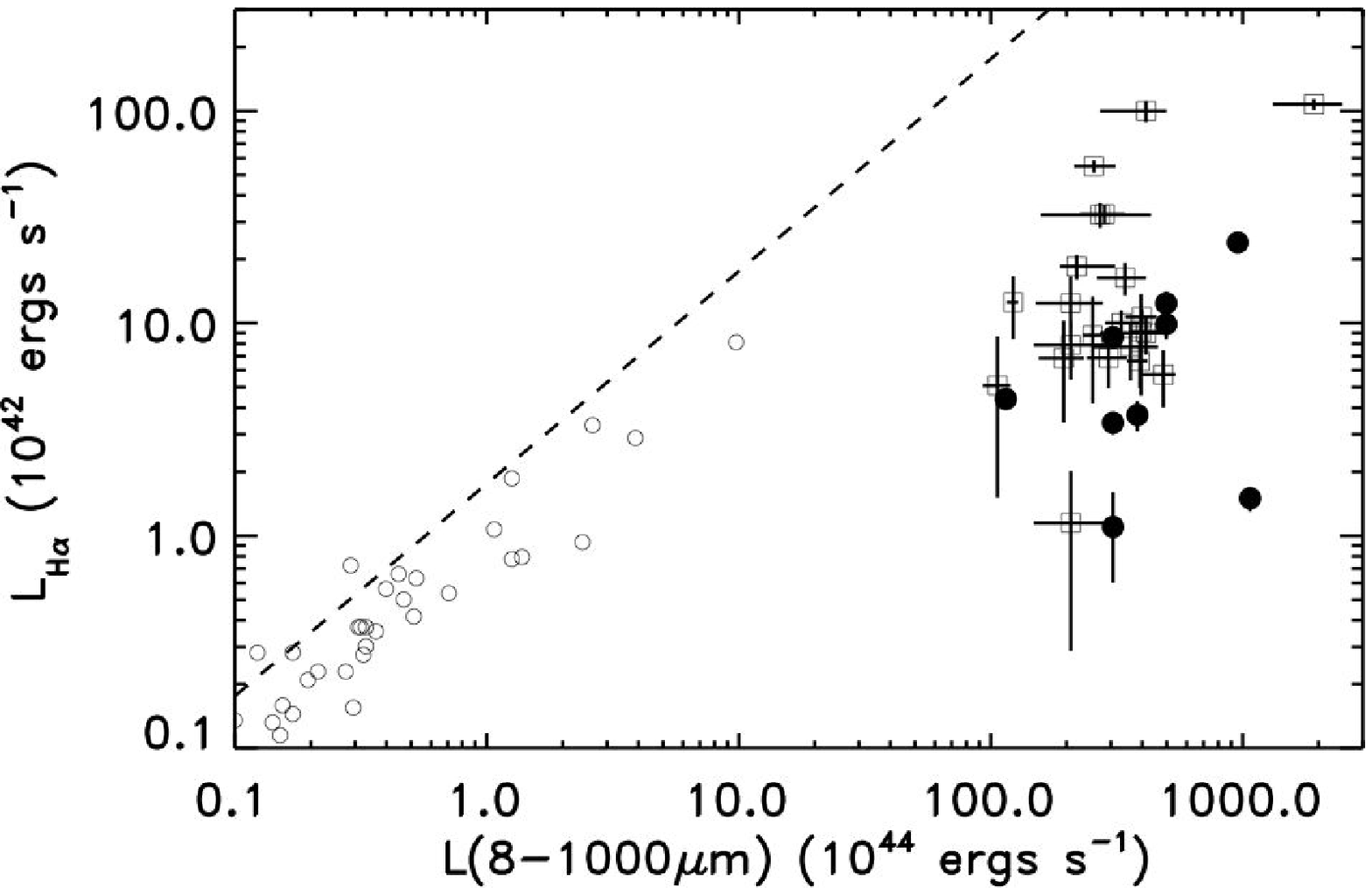}
\end{center}
{\caption[junk]{\label{fig:lums} The observed H$\alpha$ luminosity versus the infrared luminosity for the 9 sources with H$\alpha$ line measurements (large filled circles). For comparison, we plot a sample of local $IRAS$ galaxies (\citealt{kew02}; small empty circles), and sub-mm selected galaxies (\citealt{swi04}; open squares). The dotted line shows where the infrared and H$\alpha$ SFR indicators would agree. The H$\alpha$ luminosities are smaller than we would expect given the large infrared luminosities, suggesting large optical extinctions. }}
\end{figure}

In Figure~\ref{fig:lums}, we show the infrared luminosity versus the measured H$\alpha$ luminosity for our sample. Our sources have much smaller H$\alpha$ luminosities relative to their infrared luminosities when compared to local $IRAS$ galaxies \citep{kew02}. We compare our results to the sub-mm selected sample of \citet{swi04}. Their sample provides a good comparison sample given the similar redshift range and infrared luminosities. Their observations were also taken with the KECK / NIRSPEC 0.76\arcsec\ slit, so slit losses should be similar. A Kolmogorov-Smirnov test \citep{pre92} shows that there is only 5\% chance that the H$\alpha$ luminosities are drawn from the same distribution. This suggests that optically faint, infrared bright galaxies tend to have higher optical extinctions than sub-mm selected galaxies. 

\subsection{Comparision to mid-IR spectra}
\label{sec:midir}

In \citet{hou05} and \citet{wee06}, we presented mid-IR spectroscopy of $R-[24]>14$ sources using $Spitzer$ IRS. From fitting of the IRS spectra with templates, \citet{hou05} conclude that AGN probably dominate the mid-IR emission in the majority of f$_{24}>$ 0.8 mJy optically faint ULIRGs. Our results are consistent with this. \citet{wee06} presented the spectra for which we had previously been unable to determine a confident redshift in \citet{hou05}. 7/10 of our NIRSPEC sample were observed with $Spitzer$ IRS as part of our larger IRS program.  Three sources show strong silicate absorption features in their IRS spectra and have IRS redshift determinations that are broadly consistent with their NIRSPEC redshifts. The four sources with no clear spectral features in their IRS mid-IR spectra have redshifts derived from their near-IR spectra of $z$=2.180, 2.263, 2.498, and 3.312 (see Table~\ref{tab:sample}). The 9.7 $\rm \mu m$ silicate absorption feature is redshifted beyond the IRS spectral window for redshifts $z \gtsimeq$ 2.5, thus naturally explaining the featureless IRS spectra for the two highest redshift sources. The IRS spectra of SST24 J143424.4+334543 at $z$=2.263 and SST24 J142939.1+353558 at $z$=2.498 do show downturns in the flux at the long wavelength edge of the IRS window, suggesting the onset of the silicate absorption. However, SST24 J142842.9+342409 at $z$=2.180 is at low enough redshift to have shown any silicate absorption feature; its absence suggests a perhaps unusually weak feature, and we discuss this further in Section~\ref{sec:midir_dis}.

We also observed two further IRS sources (SST24 J142611.3+351217 and SST24 J142850.9+353146: source 20 and 23 in \citet{wee06}) with NIRSPEC but detected no significant continuum or line emission. The $K$-band magnitudes are such that we may not be able to detect continuum emission. Assuming that the source did not fall out of the slit, it is likely that these sources are at redshifts for which none of the powerful lines fell into the $K$-band spectral coverage. SST24 J142611+351217 is predicted by \citet{wee06} to be at $z\approx$1.6 which is too low for us to have detected H$\alpha$ in the $K$-band. SST24 J142850.9+353146 has a featureless mid-IR spectrum. If it is in the redshift range 2.7$<z<$3.1 or at $z>$3.7, none of the typically strong lines would have been detectable. 

\section{Discussion}
\label{sec:dis}

\subsection{The number density of obscured AGN}

Our near-IR spectra suggest that most f$_{24}>$0.8 mJy optically faint luminous infrared galaxies harbor AGN. This is an important confirmation of our $Spitzer$ IRS results and shows that there is a significant population of AGN missing from traditional optical surveys. Even giving conservative estimates of their bolometric luminosities, these sources are comparable in luminosity to bright optically selected quasars.\citet{bro06} found that the number density of 24 $\rm \mu m$ selected, optically unobscured AGN was comparable to the number density of optically selected AGN. They find $\approx$140 f$_{24}>$1 mJy quasars with $z>1.3$ in the entire Bo\"otes field. There are $\approx$340 f$_{24}>$1 mJy sources with $R-[24]>$14. If we assume that these sources are all at $z>1.3$ and harbor luminous AGN (as is suggested by our results and their large bolometric luminosities), then they constitute an important population of AGN with space densities of at least twice that of the optically luminous type I AGN. Although this is only a crude estimate, it agrees broadly with previous estimates of the number of obscured to un-obscured AGN (e.g., \citealt{ued03}; \citealt{mar05}). 

\subsection{Obscuration by dust}

7/10 of the sample presented in this paper are broad-line (optical type I) AGN. This fraction agrees broadly with receding torus models which predict that the broad-line region is surrounded by an optically thick molecular torus (e.g., \citealt{ant93}) and that the opening angle of the torus increases with luminosity because more luminous AGN can sublimate the dust out to larger distances (\citealt{law91}; \citealt{ars05}; \citealt{sim05}).  \citet{sim05} predict the fraction of type I AGN as a function of [OIII] luminosity. Our typical intrinsic [OIII] luminosities of $\approx 10^{43}-10^{45} \rm ~ergs~s^{-1}$ imply type I fractions of 0.6-1 depending on the adopted model. On first look, the fraction of broad line sources is consistent with what we would expect for such luminous AGN. However, this population was selected to have faint optical (i.e., rest-frame UV) magnitudes. Given that all of the sources have $R\gtsimeq22$, we might expect our sample to be biased to heavily extincted quasars, and therefore only visible as type II AGN. The visibility of the broad lines suggests that the lines of sight to the broad line region are not completely obscured and hence that it is unlikely that these quasars are being viewed through the mid-plane of a dusty, completely optically thick torus. Our results are also consistent with models in which the torus is identified with the dusty, optically thick region of the wind coming off the central accretion disk (e.g., \citealt{elv06}; \citealt{eli06}). If the obscuring clouds are clumpy, this may allow one to view the central broad-line regions in some cases.  

The H$\alpha$ / H$\beta$ Balmer decrements are large for both narrow-line and broad-line AGN in our sample. This, along with the optical faintness, suggests that the AGN continuum emission, broad-line region, and narrow-line regions are heavily extincted. Although the extinction of the AGN continuum and broad-line regions only requires extinction on small scales, extinction of the narrow-line region requires extinction on much larger scales. We estimate SMBH masses of $\sim 10^{8-9} \rm M_\odot$. Given the large infrared luminosity of the AGN, the central SMBH must also be undergoing very rapid growth. If the Magorrian relation holds, we would expect the host galaxies to have large bulges with significant accompanying starbursts which should be visible in the rest-frame UV. We see much fainter optical emission than we would expect, suggesting obscuration on very large (i.e., kpc) scales. The fact that we see hints of extended but very faint optical emission in many of our sources supports this view (Dey et al. in preparation).

\citet{mar06} propose two types of obscured AGN. The first are the traditional type II AGN whose optical light is obscured by a dusty torus ("torus-obscured" AGN). The second are AGN whose optical light is obscured on much larger scales ("host-obscured" AGN). They suggest that although receding torus models predict type II to type I ratios of $\sim$1:1, larger ratios may be needed to explain the hard X-ray background ($\sim3-4:1$; \citealt{gil01}). The discrepancy could be solved by the additional population of "host-obscured" AGN which will be counted as obscured AGN in the latter case but exhibit both type-I and type-II optical spectra. As we discuss above, the selection of optically faint luminous infrared galaxies will naturally select "host-obscured" sources. Albeit with only a small sample, our near-IR spectra show that these sources indeed have the fraction of broad- to narrow-line sources that one would expect for such luminous sources, despite being too obscured to appear in optically selected samples.

\subsection{Silicate absorption and optical extinction}
\label{sec:midir_dis}

There is no obvious correlation between the silicate absorption depths in the IRS spectra and optical rest-frame properties. SST24 J143312.7+342011 exhibits deep silicate absorption in its IRS spectrum whereas SST24 J142842.9+342409 must have only shallow silicate absorption yet both have very similar rest-frame optical spectra (with prominent broad $H\alpha$ emission lines). SST24 J143312.7+342011 and SST24 J143028.5+343221 both have deep silicate absorption features yet they exhibit broad and narrow H$\alpha$ emission lines respectively. 

Are the silicate depths consistent with the observation of broad emission lines? SST24 J143312.7+342011 has broad emission lines and has relatively deep silicate absorption (A(9.7$ \rm \mu m$)$\sim$1.2). Assuming the dust models of \citet{li01} which predict a extinction ratios of A(H$_\alpha$)/A(9.7$ \rm \mu m$) $\approx$10, this corresponds to A(H$_\alpha$)$\approx$12 which should be too large for us to have detected broad H$\alpha$ for any plausible intrinsic H$\alpha$ luminosities ($L(H\alpha) \ltsimeq 10^{46} \rm ergs~s^{-1}$).  However, this apparent discrepancy also exists for some local ULIRGs. \citet{hao06} investigate the distribution of silicate absorption strengths for a wide variety of local AGN and ULIRGs. For type I AGN, the deepest silicate absorption strength is found for Mkn 231 \citep{wee05} with A(9.7$ \rm \mu m$)=0.75 which would correspond to A(H$_\alpha$)$\approx$7.5, again implausibly high. Arp 220 has deep silicate absorption and evidence for a heavily embedded nuclear source (e.g., \citealt{spo04}), yet optical emission lines are visible and lead to a liner classification. We suggest that in these cases, the dust distribution may have a complicated geometry.  Possible scenarios include a less than 100\% covering factor due to a clumpy dust distribution or scattering of the optical light. If the sources have optically thick nuclei, the situation may be further complicated by the fact that we may be seeing down to different optical depths at different wavelengths. 

SST24 J142842.9+342409 is an interesting case given its lack of any silicate absorption feature. The silicate absorption feature should be observable at an observed wavelength of $\approx$31 $\rm \mu m$ but no such feature is observed (see \citet{wee06} figure 1; source 15). This is particularly interesting given that H$\alpha$/H$\beta>15.7$, corresponding to E(B-V)$>$1.6 and suggesting large columns of attenuating dust. Could the geometry of this source result in a weaker silicate absorption feature? The presence of broad H$\alpha$ suggests that this source is a type 1 AGN in which the torus is viewed roughly face-on. Type 1 AGN are expected to show silicate emission from the hot illuminated surface of the inner torus \citep{pie92}. For more than a decade, silicate in emission in type 1 AGN remained elusive. However, they have recently been found in powerful unobscured AGN (\citealt{sie05}; \citealt{hao05}) and in the ULIRG FSC 10214+4724 \citep{tep06}. \citet{sie05} suggest that this may be a luminosity effect, with only the most luminous AGN showing silicate in emission. If this is the case, then we might expect to see silicate emission in our sample of very luminous type 1 optically faint ULIRGS. SST24 J142842.9+342409 has the second largest infrared luminosity of sources at low enough redshifts to observe the silicate absorption feature. Perhaps the combination of silicate emission from the torus and silicate absorption further out, results in only a weak silicate feature in this case.

\section{Conclusions}

We have obtained near-infrared spectroscopy for a small sample of 24 $\mu$m sources with extremely red optical-to-mid-infrared colors selected from the {\it Spitzer} MIPS survey of the NDWFS Bo\"otes field. We conclude that:
 
 \begin{itemize}
\item[(1)]{All of the sources in the sample lie at high redshift ($1.3<z<3.4$) and are therefore very luminous infrared sources.}
\item[(2)]{All the sources show either broad-lines or have narrow-line ratios suggesting that they contain powerful AGN. This is an important confirmation of our $Spitzer$ IRS results and shows that there is a significant population of AGN missing from traditional optical surveys. If all $R-[24]>14$ sources with f$_{24}>$ 1 mJy harbor AGN, then these sources may be twice as common as more optically luminous type I AGN at similar redshifts.}
\item[(3)]{7/10 of the sample have broad ($>1900\rm~km~s^{-1}$) H$\alpha$ or H$\beta$ emission lines. We classify these sources as type I AGN. The remaining 3 sources have Balmer line widths $>450\rm~km~s^{-1}$ and line diagnostics predict them to be type II AGN. }
\item[(4)]{The fraction of type I to type II AGN is consistent with receding torus models given their large intrinsic [OIII] luminosities. This is somewhat surprising: given their high infrared luminosities and faint optical magnitudes, we might expect these sources to all be heavily extincted quasars (i.e., optical type II sources).}
\item[(5)]{Our limits on the H$\alpha$/H$\beta$ ratio in 4 cases suggests that the optical emission in both narrow- and broad-line optically faint ULIRGs is heavily extincted (by at least 2.4 magnitudes at the wavelength of H$\alpha$). Since the narrow-line region is also extincted, this suggests dust obscuration on large scales.}
\item[(6)]{Given that these sources are bolometrically powerful, we would expect the AGN to be hosted in a large galaxy which is actively forming stars and optically bright. However, the UV continuum emission from the host galaxies is faint and in some cases may be extended. We suggest that these sources are examples of 'host obscured' AGN in which a large proportion of the obscuration is contributed on large ($\sim$ kpc) scales within the host galaxies.\\}
 \end{itemize}
 
We intend to obtain similar observations of a larger sample of optically faint, luminous infrared galaxies to determine the fraction of type I to type II AGN more accurately and to get better constraints on dust extinction within these sources.

We thank our colleagues on the NDWFS, MIPS, IRS, and XBo\"otes teams. KB acknowledges the support provided by the Giacconi fellowship. AD and BTJ are supported by the National Optical Astronomy Observatory, which is operated by the Association of Universities for Research in Astronomy, Inc. (AURA) under cooperative agreement with the National Science Foundation. This work is based in part on observations made with the {\it Spitzer Space Telescope}, which is operated by the Jet Propulsion Laboratory, California Institute of Technology under a contract with NASA. Support for this work was provided by NASA through awards issued by JPL/Caltech. The {\it Spitzer MIPS} survey of the Bo\"otes region was obtained using GTO time provided by the {\it Spitzer} Infrared Spectrograph Team (James Houck, P.I.) and by M. Rieke. Some of the data presented herein were obtained at the W.M. Keck Observatory, which is operated as a scientific partnership between the California Institute of Technology, the University of California and the National Aeronautics and Space Administration (NASA). It was made possible by the generous financial support of the W.M. Keck Foundation. We thank the staff of the W.M. Keck Observatory for their assistance in making these observations possible. We extend special thanks to those of Hawaiian ancestry on whose sacred mountain we are privileged to be guests.  Without their generous hospitality, none of the observations presented herein would have been possible. The paper is also based in part on observations obtained at the Gemini Observatory, which is operated by the Association of Universities for Research in Astronomy, Inc., under a cooperative agreement with the NSF on behalf of the Gemini partnership: the National Science Foundation (United States), the Particle Physics and Astronomy Research Council (United Kingdom), the National Research Council (Canada), CONICYT (Chile), the Australian Research Council (Australia), CNPq (Brazil) and CONICET (Argentina).  The authors wish to recognize and acknowledge the very significant cultural role and reverence that the summit of Mauna Kea has always had within the indigenous Hawaiian community.  We are most fortunate to have the opportunity to conduct observations from this mountain. We thank the anonymous referee for his/her useful comments.

\bibliography{ms}

\begin{thebibliography}{57}
\expandafter\ifx\csname natexlab\endcsname\relax\def\natexlab#1{#1}\fi

\bibitem[{{Antonucci}(1993)}]{ant93}
{Antonucci}, R. 1993, \araa, 31, 473

\bibitem[{{Arshakian}(2005)}]{ars05}
{Arshakian}, T.~G. 2005, \aap, 436, 817

\bibitem[{{Baldwin} {et~al.}(1981){Baldwin}, {Phillips}, \&
  {Terlevich}}]{bal81}
{Baldwin}, J.~A., {Phillips}, M.~M., \& {Terlevich}, R. 1981, \pasp, 93, 5

\bibitem[{{Barger} {et~al.}(2001){Barger}, {Cowie}, {Bautz}, {Brandt},
  {Garmire}, {Hornschemeier}, {Ivison}, \& {Owen}}]{bar01}
{Barger}, A.~J., {Cowie}, L.~L., {Bautz}, M.~W., {Brandt}, W.~N., {Garmire},
  G.~P., {Hornschemeier}, A.~E., {Ivison}, R.~J., \& {Owen}, F.~N. 2001, \aj,
  122, 2177

\bibitem[{{Brand} {et~al.}(2006{\natexlab{a}}){Brand}, {Brown}, {Dey},
  {Jannuzi}, {Kochanek}, {Kenter}, {Fabricant}, {Fazio}, {Forman}, {Green},
  {Jones}, {McNamara}, {Murray}, {Najita}, {Rieke}, {Shields}, \&
  {Vikhlinin}}]{bra06b}
{Brand}, K., {Brown}, M.~J.~I., {Dey}, A., {Jannuzi}, B.~T., {Kochanek}, C.~S.,
  {Kenter}, A.~T., {Fabricant}, D., {Fazio}, G.~G., {Forman}, W.~R., {Green},
  P.~J., {Jones}, C.~J., {McNamara}, B.~R., {Murray}, S.~S., {Najita}, J.~R.,
  {Rieke}, M., {Shields}, J.~C., \& {Vikhlinin}, A. 2006{\natexlab{a}}, \apj,
  641, 140

\bibitem[{{Brand} {et~al.}(2006{\natexlab{b}}){Brand}, {Dey}, {Weedman},
  {Desai}, {Le Floc'h}, {Jannuzi}, {Soifer}, {Brown}, {Eisenhardt}, {Gorjian},
  {Papovich}, {Smith}, {Willner}, \& {Cool}}]{bra06}
{Brand}, K., {Dey}, A., {Weedman}, D., {Desai}, V., {Le Floc'h}, E., {Jannuzi},
  B.~T., {Soifer}, B.~T., {Brown}, M.~J.~I., {Eisenhardt}, P., {Gorjian}, V.,
  {Papovich}, C., {Smith}, H.~A., {Willner}, S.~P., \& {Cool}, R.~J.
  2006{\natexlab{b}}, \apj, 644, 143

\bibitem[{{Brown} {et~al.}(2006){Brown}, {Brand}, {Dey}, {Jannuzi}, {Cool}, {Le
  Floc'h}, {Kochanek}, {Armus}, {Bian}, {Higdon}, {Higdon}, {Papovich},
  {Rieke}, {Rieke}, {Smith}, {Soifer}, \& {Weedman}}]{bro06}
{Brown}, M.~J.~I., {Brand}, K., {Dey}, A., {Jannuzi}, B.~T., {Cool}, R., {Le
  Floc'h}, E., {Kochanek}, C.~S., {Armus}, L., {Bian}, C., {Higdon}, J.,
  {Higdon}, S., {Papovich}, C., {Rieke}, G., {Rieke}, M., {Smith}, J.~D.,
  {Soifer}, B.~T., \& {Weedman}, D. 2006, \apj, 638, 88

\bibitem[{{Desai} {et~al.}(2006){Desai}, {Armus}, {Soifer}, {Weedman},
  {Higdon}, {Bian}, {Borys}, {Spoon}, {Charmandaris}, {Brand}, {Brown}, {Dey},
  {Higdon}, {Houck}, {Jannuzi}, {Le Floc'h}, {Ashby}, \& {Smith}}]{des06}
{Desai}, V., {Armus}, L., {Soifer}, B.~T., {Weedman}, D.~W., {Higdon}, S.,
  {Bian}, C., {Borys}, C., {Spoon}, H.~W.~W., {Charmandaris}, V., {Brand}, K.,
  {Brown}, M.~J.~I., {Dey}, A., {Higdon}, J., {Houck}, J., {Jannuzi}, B.~T.,
  {Le Floc'h}, E., {Ashby}, M.~L.~N., \& {Smith}, H.~A. 2006, \apj, 641, 133

\bibitem[{{Dudley}(1999)}]{dud99}
{Dudley}, C.~C. 1999, \mnras, 307, 553

\bibitem[{{Elitzur} \& {Shlosman}(2006)}]{eli06}
{Elitzur}, M., \& {Shlosman}, I. 2006, \apjl, 648, L101

\bibitem[{{Elston} {et~al.}(2006){Elston}, {Gonzalez}, {McKenzie}, {Brodwin},
  {Brown}, {Cardona}, {Dey}, {Dickinson}, {Eisenhardt}, {Jannuzi}, {Lin},
  {Mohr}, {Raines}, {Stanford}, \& {Stern}}]{els06}
{Elston}, R.~J., {Gonzalez}, A.~H., {McKenzie}, E., {Brodwin}, M., {Brown},
  M.~J.~I., {Cardona}, G., {Dey}, A., {Dickinson}, M., {Eisenhardt}, P.~R.,
  {Jannuzi}, B.~T., {Lin}, Y.-T., {Mohr}, J.~J., {Raines}, S.~N., {Stanford},
  S.~A., \& {Stern}, D. 2006, \apj, 639, 816

\bibitem[{{Elvis}(2006)}]{elv06}
{Elvis}, M. 2006, Memorie della Societa Astronomica Italiana, 77, 573

\bibitem[{{Erb} {et~al.}(2006){Erb}, {Steidel}, {Shapley}, {Pettini}, {Reddy},
  \& {Adelberger}}]{erb06}
{Erb}, D.~K., {Steidel}, C.~C., {Shapley}, A.~E., {Pettini}, M., {Reddy},
  N.~A., \& {Adelberger}, K.~L. 2006, \apj, 647, 128

\bibitem[{{Gaskell} \& {Ferland}(1984)}]{gas84}
{Gaskell}, C.~M., \& {Ferland}, G.~J. 1984, \pasp, 96, 393

\bibitem[{{Gilli} {et~al.}(2001){Gilli}, {Salvati}, \& {Hasinger}}]{gil01}
{Gilli}, R., {Salvati}, M., \& {Hasinger}, G. 2001, \aap, 366, 407

\bibitem[{{Greene} \& {Ho}(2005)}]{gre05}
{Greene}, J.~E., \& {Ho}, L.~C. 2005, \apj, 630, 122

\bibitem[{{Hao} {et~al.}(2006){Hao}, {Mao}, {Deng}, {Xia}, \& {Wu}}]{hao06}
{Hao}, C.~N., {Mao}, S., {Deng}, Z.~G., {Xia}, X.~Y., \& {Wu}, H. 2006, \mnras,
  373, 1264

\bibitem[{{Hao} {et~al.}(2005){Hao}, {Spoon}, {Sloan}, {Marshall}, {Armus},
  {Tielens}, {Sargent}, {van Bemmel}, {Charmandaris}, {Weedman}, \&
  {Houck}}]{hao05}
{Hao}, L., {Spoon}, H.~W.~W., {Sloan}, G.~C., {Marshall}, J.~A., {Armus}, L.,
  {Tielens}, A.~G.~G.~M., {Sargent}, B., {van Bemmel}, I.~M., {Charmandaris},
  V., {Weedman}, D.~W., \& {Houck}, J.~R. 2005, \apjl, 625, L75

\bibitem[{{Hodapp} {et~al.}(2003){Hodapp}, {Jensen}, {Irwin}, {Yamada},
  {Chung}, {Fletcher}, {Robertson}, {Hora}, {Simons}, {Mays}, {Nolan}, {Bec},
  {Merrill}, \& {Fowler}}]{hod03}
{Hodapp}, K.~W., {Jensen}, J.~B., {Irwin}, E.~M., {Yamada}, H., {Chung}, R.,
  {Fletcher}, K., {Robertson}, L., {Hora}, J.~L., {Simons}, D.~A., {Mays}, W.,
  {Nolan}, R., {Bec}, M., {Merrill}, M., \& {Fowler}, A.~M. 2003, \pasp, 115,
  1388

\bibitem[{{Houck} {et~al.}(2005){Houck}, {Soifer}, {Weedman}, {Higdon},
  {Higdon}, {Herter}, {Brown}, {Dey}, {Jannuzi}, {Le Floc'h}, {Rieke}, {Armus},
  {Charmandaris}, {Brandl}, \& {Teplitz}}]{hou05}
{Houck}, J.~R., {Soifer}, B.~T., {Weedman}, D., {Higdon}, S.~J.~U., {Higdon},
  J.~L., {Herter}, T., {Brown}, M.~J.~I., {Dey}, A., {Jannuzi}, B.~T., {Le
  Floc'h}, E., {Rieke}, M., {Armus}, L., {Charmandaris}, V., {Brandl}, B.~R.,
  \& {Teplitz}, H.~I. 2005, \apjl, 622, L105

\bibitem[{{Jannuzi} \& {Dey}(1999)}]{jan99}
{Jannuzi}, B.~T., \& {Dey}, A. 1999, in ASP Conf. Ser. 191: Photometric
  Redshifts and the Detection of High Redshift Galaxies, ed. R.~{Weymann},
  L.~{Storrie-Lombardi}, M.~{Sawicki}, \& R.~{Brunner}, 111--+

\bibitem[{{Kauffmann} {et~al.}(2003){Kauffmann}, {Heckman}, {Tremonti},
  {Brinchmann}, {Charlot}, {White}, {Ridgway}, {Brinkmann}, {Fukugita}, {Hall},
  {Ivezi{\'c}}, {Richards}, \& {Schneider}}]{kau03}
{Kauffmann}, G., {Heckman}, T.~M., {Tremonti}, C., {Brinchmann}, J., {Charlot},
  S., {White}, S.~D.~M., {Ridgway}, S.~E., {Brinkmann}, J., {Fukugita}, M.,
  {Hall}, P.~B., {Ivezi{\'c}}, {\v Z}., {Richards}, G.~T., \& {Schneider},
  D.~P. 2003, \mnras, 346, 1055

\bibitem[{{Kenter} {et~al.}(2005){Kenter}, {Murray}, {Forman}, {Jones},
  {Green}, {Kochanek}, {Vikhlinin}, {Fabricant}, {Fazio}, {Brand}, {Brown},
  {Dey}, {Jannuzi}, {Najita}, {McNamara}, {Shields}, \& {Rieke}}]{ken05}
{Kenter}, A., {Murray}, S.~S., {Forman}, W.~R., {Jones}, C., {Green}, P.,
  {Kochanek}, C.~S., {Vikhlinin}, A., {Fabricant}, D., {Fazio}, G., {Brand},
  K., {Brown}, M.~J.~I., {Dey}, A., {Jannuzi}, B.~T., {Najita}, J., {McNamara},
  B., {Shields}, J., \& {Rieke}, M. 2005, \apjs, 161, 9

\bibitem[{{Kewley} {et~al.}(2001){Kewley}, {Dopita}, {Sutherland}, {Heisler},
  \& {Trevena}}]{kew01}
{Kewley}, L.~J., {Dopita}, M.~A., {Sutherland}, R.~S., {Heisler}, C.~A., \&
  {Trevena}, J. 2001, \apj, 556, 121

\bibitem[{{Kewley} {et~al.}(2002){Kewley}, {Geller}, {Jansen}, \&
  {Dopita}}]{kew02}
{Kewley}, L.~J., {Geller}, M.~J., {Jansen}, R.~A., \& {Dopita}, M.~A. 2002,
  \aj, 124, 3135

\bibitem[{{Kewley} {et~al.}(2006){Kewley}, {Groves}, {Kauffmann}, \&
  {Heckman}}]{kew06}
{Kewley}, L.~J., {Groves}, B., {Kauffmann}, G., \& {Heckman}, T. 2006, \mnras,
  372, 961

\bibitem[{{Lawrence}(1991)}]{law91}
{Lawrence}, A. 1991, \mnras, 252, 586

\bibitem[{{Le Floc'h} {et~al.}(2004){Le Floc'h}, {P{\'e}rez-Gonz{\'a}lez},
  {Rieke}, {Papovich}, {Huang}, {Barmby}, {Dole}, {Egami}, {Alonso-Herrero},
  {Wilson}, {Miyazaki}, {Rigby}, {Bei}, {Blaylock}, {Engelbracht}, {Fazio},
  {Frayer}, {Gordon}, {Hines}, {Misselt}, {Morrison}, {Muzerolle}, {Rieke},
  {Rigopoulou}, {Su}, {Willner}, \& {Young}}]{lef04}
{Le Floc'h}, E., {P{\'e}rez-Gonz{\'a}lez}, P.~G., {Rieke}, G.~H., {Papovich},
  C., {Huang}, J.-S., {Barmby}, P., {Dole}, H., {Egami}, E., {Alonso-Herrero},
  A., {Wilson}, G., {Miyazaki}, S., {Rigby}, J.~R., {Bei}, L., {Blaylock}, M.,
  {Engelbracht}, C.~W., {Fazio}, G.~G., {Frayer}, D.~T., {Gordon}, K.~D.,
  {Hines}, D.~C., {Misselt}, K.~A., {Morrison}, J.~E., {Muzerolle}, J.,
  {Rieke}, M.~J., {Rigopoulou}, D., {Su}, K.~Y.~L., {Willner}, S.~P., \&
  {Young}, E.~T. 2004, \apjs, 154, 170

\bibitem[{{Levenson} {et~al.}(2006){Levenson}, {Sirocky}, {Hao}, {Spoon},
  {Marshall}, {Elitzur}, \& {Houck}}]{lev06}
{Levenson}, N.~A., {Sirocky}, M.~M., {Hao}, L., {Spoon}, H.~W.~W., {Marshall},
  J.~A., {Elitzur}, M., \& {Houck}, J.~R. 2006, ArXiv Astrophysics e-prints

\bibitem[{{Li} \& {Draine}(2001)}]{li01}
{Li}, A., \& {Draine}, B.~T. 2001, \apj, 554, 778

\bibitem[{{Mart{\'{\i}}nez-Sansigre} {et~al.}(2006){Mart{\'{\i}}nez-Sansigre},
  {Rawlings}, {Lacy}, {Fadda}, {Jarvis}, {Marleau}, {Simpson}, \&
  {Willott}}]{mar06}
{Mart{\'{\i}}nez-Sansigre}, A., {Rawlings}, S., {Lacy}, M., {Fadda}, D.,
  {Jarvis}, M.~J., {Marleau}, F.~R., {Simpson}, C., \& {Willott}, C.~J. 2006,
  \mnras, 370, 1479

\bibitem[{{Mart{\'{\i}}nez-Sansigre} {et~al.}(2005){Mart{\'{\i}}nez-Sansigre},
  {Rawlings}, {Lacy}, {Fadda}, {Marleau}, {Simpson}, {Willott}, \&
  {Jarvis}}]{mar05}
{Mart{\'{\i}}nez-Sansigre}, A., {Rawlings}, S., {Lacy}, M., {Fadda}, D.,
  {Marleau}, F.~R., {Simpson}, C., {Willott}, C.~J., \& {Jarvis}, M.~J. 2005,
  \nat, 436, 666

\bibitem[{{McLean} {et~al.}(1998){McLean}, {Becklin}, {Bendiksen}, {Brims},
  {Canfield}, {Figer}, {Graham}, {Hare}, {Lacayanga}, {Larkin}, {Larson},
  {Levenson}, {Magnone}, {Teplitz}, \& {Wong}}]{mcl98}
{McLean}, I.~S., {Becklin}, E.~E., {Bendiksen}, O., {Brims}, G., {Canfield},
  J., {Figer}, D.~F., {Graham}, J.~R., {Hare}, J., {Lacayanga}, F., {Larkin},
  J.~E., {Larson}, S.~B., {Levenson}, N., {Magnone}, N., {Teplitz}, H., \&
  {Wong}, W. 1998, in Proc. SPIE Vol. 3354, p. 566-578, Infrared Astronomical
  Instrumentation, Albert M. Fowler; Ed., ed. A.~M. {Fowler}, 566--578

\bibitem[{{Murray} {et~al.}(2005){Murray}, {Kenter}, {Forman}, {Jones},
  {Green}, {Kochanek}, {Vikhlinin}, {Fabricant}, {Fazio}, {Brand}, {Brown},
  {Dey}, {Jannuzi}, {Najita}, {McNamara}, {Shields}, \& {Rieke}}]{mur05}
{Murray}, S.~S., {Kenter}, A., {Forman}, W.~R., {Jones}, C., {Green}, P.~J.,
  {Kochanek}, C.~S., {Vikhlinin}, A., {Fabricant}, D., {Fazio}, G., {Brand},
  K., {Brown}, M.~J.~I., {Dey}, A., {Jannuzi}, B.~T., {Najita}, J., {McNamara},
  B., {Shields}, J., \& {Rieke}, M. 2005, \apjs, 161, 1

\bibitem[{{Osterbrock}(1989)}]{ost89}
{Osterbrock}, D.~E. 1989, {Astrophysics of gaseous nebulae and active galactic
  nuclei} (Research supported by the University of California, John Simon
  Guggenheim Memorial Foundation, University of Minnesota, et al.~Mill Valley,
  CA, University Science Books, 1989, 422 p.)

\bibitem[{{Osterbrock} \& {Pogge}(1985)}]{ost85}
{Osterbrock}, D.~E., \& {Pogge}, R.~W. 1985, \apj, 297, 166

\bibitem[{{Pier} \& {Krolik}(1992)}]{pie92}
{Pier}, E.~A., \& {Krolik}, J.~H. 1992, \apj, 401, 99

\bibitem[{{Poggianti} \& {Wu}(2000)}]{pog00}
{Poggianti}, B.~M., \& {Wu}, H. 2000, \apj, 529, 157

\bibitem[{{Press} {et~al.}(1992){Press}, {Teukolsky}, {Vetterling}, \&
  {Flannery}}]{pre92}
{Press}, W.~H., {Teukolsky}, S.~A., {Vetterling}, W.~T., \& {Flannery}, B.~P.
  1992, {Numerical recipes in C. The art of scientific computing} (Cambridge:
  University Press, |c1992, 2nd ed.)

\bibitem[{{Richards} {et~al.}(2003){Richards}, {Hall}, {Vanden Berk},
  {Strauss}, {Schneider}, {Weinstein}, {Reichard}, {York}, {Knapp}, {Fan},
  {Ivezi{\'c}}, {Brinkmann}, {Budav{\'a}ri}, {Csabai}, \& {Nichol}}]{ric03}
{Richards}, G.~T., {Hall}, P.~B., {Vanden Berk}, D.~E., {Strauss}, M.~A.,
  {Schneider}, D.~P., {Weinstein}, M.~A., {Reichard}, T.~A., {York}, D.~G.,
  {Knapp}, G.~R., {Fan}, X., {Ivezi{\'c}}, {\v Z}., {Brinkmann}, J.,
  {Budav{\'a}ri}, T., {Csabai}, I., \& {Nichol}, R.~C. 2003, \aj, 126, 1131

\bibitem[{{Siebenmorgen} {et~al.}(2005){Siebenmorgen}, {Haas}, {Kruegel}, \&
  {Schulz}}]{sie05}
{Siebenmorgen}, R., {Haas}, M., {Kruegel}, E., \& {Schulz}, B. 2005,
  Astronomische Nachrichten, 326, 556

\bibitem[{{Simpson}(2005)}]{sim05}
{Simpson}, C. 2005, \mnras, 360, 565

\bibitem[{{Spoon} {et~al.}(2006){Spoon}, {Marshall}, {Houck}, {Elitzur}, {Hao},
  {Armus}, {Brandl}, \& {Charmandaris}}]{spo06}
{Spoon}, H.~W.~W., {Marshall}, J.~A., {Houck}, J.~R., {Elitzur}, M., {Hao}, L.,
  {Armus}, L., {Brandl}, B.~R., \& {Charmandaris}, V. 2006, ArXiv Astrophysics
  e-prints

\bibitem[{{Spoon} {et~al.}(2004){Spoon}, {Moorwood}, {Lutz}, {Tielens},
  {Siebenmorgen}, \& {Keane}}]{spo04}
{Spoon}, H.~W.~W., {Moorwood}, A.~F.~M., {Lutz}, D., {Tielens}, A.~G.~G.~M.,
  {Siebenmorgen}, R., \& {Keane}, J.~V. 2004, \aap, 414, 873

\bibitem[{{Storey} \& {Zeippen}(2000)}]{sto00}
{Storey}, P.~J., \& {Zeippen}, C.~J. 2000, \mnras, 312, 813

\bibitem[{{Swinbank} {et~al.}(2004){Swinbank}, {Smail}, {Chapman}, {Blain},
  {Ivison}, \& {Keel}}]{swi04}
{Swinbank}, A.~M., {Smail}, I., {Chapman}, S.~C., {Blain}, A.~W., {Ivison},
  R.~J., \& {Keel}, W.~C. 2004, \apj, 617, 64

\bibitem[{{Takata} {et~al.}(2006){Takata}, {Sekiguchi}, {Smail}, {Chapman},
  {Geach}, {Swinbank}, {Blain}, \& {Ivison}}]{tak06}
{Takata}, T., {Sekiguchi}, K., {Smail}, I., {Chapman}, S.~C., {Geach}, J.~E.,
  {Swinbank}, A.~M., {Blain}, A., \& {Ivison}, R.~J. 2006, \apj, 651, 713

\bibitem[{{Teplitz} {et~al.}(2006){Teplitz}, {Armus}, {Soifer}, {Charmandaris},
  {Marshall}, {Spoon}, {Lawrence}, {Hao}, {Higdon}, {Wu}, {Lacy}, {Eisenhardt},
  {Herter}, \& {Houck}}]{tep06}
{Teplitz}, H.~I., {Armus}, L., {Soifer}, B.~T., {Charmandaris}, V., {Marshall},
  J.~A., {Spoon}, H., {Lawrence}, C., {Hao}, L., {Higdon}, S., {Wu}, Y.,
  {Lacy}, M., {Eisenhardt}, P.~R., {Herter}, T., \& {Houck}, J.~R. 2006, \apjl,
  638, L1

\bibitem[{{Ueda} {et~al.}(2003){Ueda}, {Akiyama}, {Ohta}, \& {Miyaji}}]{ued03}
{Ueda}, Y., {Akiyama}, M., {Ohta}, K., \& {Miyaji}, T. 2003, \apj, 598, 886

\bibitem[{{van Dokkum} {et~al.}(2004){van Dokkum}, {Franx}, {F{\"o}rster
  Schreiber}, {Illingworth}, {Daddi}, {Knudsen}, {Labb{\'e}}, {Moorwood},
  {Rix}, {R{\"o}ttgering}, {Rudnick}, {Trujillo}, {van der Werf}, {van der
  Wel}, {van Starkenburg}, \& {Wuyts}}]{van04}
{van Dokkum}, P.~G., {Franx}, M., {F{\"o}rster Schreiber}, N.~M.,
  {Illingworth}, G.~D., {Daddi}, E., {Knudsen}, K.~K., {Labb{\'e}}, I.,
  {Moorwood}, A., {Rix}, H.-W., {R{\"o}ttgering}, H., {Rudnick}, G.,
  {Trujillo}, I., {van der Werf}, P., {van der Wel}, A., {van Starkenburg}, L.,
  \& {Wuyts}, S. 2004, \apj, 611, 703

\bibitem[{{Veilleux} {et~al.}(1995){Veilleux}, {Kim}, {Sanders}, {Mazzarella},
  \& {Soifer}}]{vei95}
{Veilleux}, S., {Kim}, D.-C., {Sanders}, D.~B., {Mazzarella}, J.~M., \&
  {Soifer}, B.~T. 1995, \apjs, 98, 171

\bibitem[{{Veilleux} \& {Osterbrock}(1987)}]{vei87}
{Veilleux}, S., \& {Osterbrock}, D.~E. 1987, \apjs, 63, 295

\bibitem[{{Weedman} {et~al.}(2005){Weedman}, {Hao}, {Higdon}, {Devost}, {Wu},
  {Charmandaris}, {Brandl}, {Bass}, \& {Houck}}]{wee05}
{Weedman}, D.~W., {Hao}, L., {Higdon}, S.~J.~U., {Devost}, D., {Wu}, Y.,
  {Charmandaris}, V., {Brandl}, B., {Bass}, E., \& {Houck}, J.~R. 2005, \apj,
  633, 706

\bibitem[{{Weedman} {et~al.}(2006){Weedman}, {Soifer}, {Hao}, {Higdon},
  {Higdon}, {Houck}, {Le Floc'h}, {Brown}, {Dey}, {Jannuzi}, {Rieke}, {Desai},
  {Bian}, {Thompson}, {Armus}, {Teplitz}, {Eisenhardt}, \& {Willner}}]{wee06}
{Weedman}, D.~W., {Soifer}, B.~T., {Hao}, L., {Higdon}, J.~L., {Higdon},
  S.~J.~U., {Houck}, J.~R., {Le Floc'h}, E., {Brown}, M.~J.~I., {Dey}, A.,
  {Jannuzi}, B.~T., {Rieke}, M., {Desai}, V., {Bian}, C., {Thompson}, D.,
  {Armus}, L., {Teplitz}, H., {Eisenhardt}, P., \& {Willner}, S.~P. 2006, \apj,
  651, 101

\bibitem[{{Werner} {et~al.}(2004){Werner}, {Roellig}, {Low}, {Rieke}, {Rieke},
  {Hoffmann}, {Young}, {Houck}, {Brandl}, {Fazio}, {Hora}, {Gehrz}, {Helou},
  {Soifer}, {Stauffer}, {Keene}, {Eisenhardt}, {Gallagher}, {Gautier}, {Irace},
  {Lawrence}, {Simmons}, {Van Cleve}, {Jura}, {Wright}, \&
  {Cruikshank}}]{wer04}
{Werner}, M.~W., {Roellig}, T.~L., {Low}, F.~J., {Rieke}, G.~H., {Rieke}, M.,
  {Hoffmann}, W.~F., {Young}, E., {Houck}, J.~R., {Brandl}, B., {Fazio}, G.~G.,
  {Hora}, J.~L., {Gehrz}, R.~D., {Helou}, G., {Soifer}, B.~T., {Stauffer}, J.,
  {Keene}, J., {Eisenhardt}, P., {Gallagher}, D., {Gautier}, T.~N., {Irace},
  W., {Lawrence}, C.~R., {Simmons}, L., {Van Cleve}, J.~E., {Jura}, M.,
  {Wright}, E.~L., \& {Cruikshank}, D.~P. 2004, \apjs, 154, 1

\bibitem[{{Xu} {et~al.}(2001){Xu}, {Lonsdale}, {Shupe}, {O'Linger}, \&
  {Masci}}]{xu01}
{Xu}, C., {Lonsdale}, C.~J., {Shupe}, D.~L., {O'Linger}, J., \& {Masci}, F.
  2001, \apj, 562, 179

\bibitem[{{Yan} {et~al.}(2005){Yan}, {Chary}, {Armus}, {Teplitz}, {Helou},
  {Frayer}, {Fadda}, {Surace}, \& {Choi}}]{yan05}
{Yan}, L., {Chary}, R., {Armus}, L., {Teplitz}, H., {Helou}, G., {Frayer}, D.,
  {Fadda}, D., {Surace}, J., \& {Choi}, P. 2005, \apj, 628, 604

\end{thebibliography}

\end{document}